\documentclass[12pt,psfig,epsf,bbox]{article}

\begin{document}

\newcommand{\vAi}{{\cal A}_{i_1\cdots i_n}} \newcommand{\vAim}{{\cal
A}_{i_1\cdots i_{n-1}}} \newcommand{\vAbi}{\bar{\cal A}^{i_1\cdots i_n}}
\newcommand{\vAbim}{\bar{\cal A}^{i_1\cdots i_{n-1}}}
\newcommand{\htS}{\hat{S}} \newcommand{\htR}{\hat{R}}
\newcommand{\htB}{\hat{B}} \newcommand{\htD}{\hat{D}}
\newcommand{\htV}{\hat{V}} \newcommand{\cT}{{\cal T}} \newcommand{\cM}{{\cal
M}} \newcommand{\cMs}{{\cal M}^*}
 \newcommand{\vk}{{\bf k}}
\newcommand{\vK}{{\vec K}} \newcommand{\vb}{{\vec b}} \newcommand{{\vp}}{{\vec
p}} \newcommand{{\vq}}{{\vec q}} \newcommand{\vQ}{{\vec Q}}
\newcommand{\vx}{{\vec x}}
\newcommand{\tr}{{{\rm Tr}}} 
\newcommand{\beq}{\begin{equation}}
\newcommand{\eeq}[1]{\label{#1} \end{equation}} 
\newcommand{\half}{{\textstyle
\frac{1}{2}}} \newcommand{\gton}{\stackrel{>}{\sim}}
\newcommand{\lton}{\mathrel{\lower.9ex \hbox{$\stackrel{\displaystyle
<}{\sim}$}}} \newcommand{\ee}{\end{equation}}
\newcommand{\ben}{\begin{enumerate}} \newcommand{\een}{\end{enumerate}}
\newcommand{\bit}{\begin{itemize}} \newcommand{\eit}{\end{itemize}}
\newcommand{\bc}{\begin{center}} \newcommand{\ec}{\end{center}}
\newcommand{\bea}{\begin{eqnarray}} \newcommand{\eea}{\end{eqnarray}}
\newcommand{\beqar}{\begin{eqnarray}} \newcommand{\eeqar}[1]{\label{#1}
\end{eqnarray}} \newcommand{\bra}[1]{\langle {#1}|}
\newcommand{\ket}[1]{|{#1}\rangle}
\newcommand{\norm}[2]{\langle{#1}|{#2}\rangle}
\newcommand{\brac}[3]{\langle{#1}|{#2}|{#3}\rangle} \newcommand{\hilb}{{\cal
H}} \newcommand{\pleft}{\stackrel{\leftarrow}{\partial}}
\newcommand{\pright}{\stackrel{\rightarrow}{\partial}}

\begin{center}
{\Large {\bf{Quantitative modeling and data analysis of SELEX experiments}}}

\vspace{1cm}

{Marko Djordjevic,$^{1,2,*}$ and Anirvan M. Sengupta$^3$ }

\vspace{.8cm}

$^{1}${\em{Department of Physics, Columbia University,  
           New York, NY 10027 }} 

$^{2}${\em{Mathematical Biosciences Institute, The Ohio State University,  
           Columbus, OH 43210 }} 

$^{3}${\em{Department of Physics and BioMaPS Institute, 
           Rutgers University, Piscataway, NJ 08854}} 

\vspace{.4cm}
$^{*}${\em{Corresponding author: Marko Djordjevic, Mathematical Biosciences 
           Institute, The The Ohio State University,  Columbus, OH 43210
           Phone: (614) 292-6159, FAX: (614) 247-6643; \\ 
           E-mail: mdjordjevic@math.ohio-state.edu}} 

\vspace{.5cm}

\today
\end{center}

\vspace{.5cm}

\begin{abstract}
SELEX (Systematic Evolution of Ligands by Exponential Enrichment) is an 
experimental procedure that allows extracting, from an initially random pool 
of DNA, those oligomers with high affinity for a given DNA-binding protein. We 
address what is a suitable experimental and computational procedure to infer 
parameters of transcription factor-DNA interaction from SELEX experiments. To 
answer this, we use a biophysical model of transcription factor-DNA 
interactions to quantitatively model SELEX. We show that a standard procedure 
is unsuitable for obtaining accurate interaction parameters. However, we 
theoretically show that a modified experiment in which chemical potential is 
fixed through different rounds of the experiment allows robust generation of 
an appropriate data set. Based on our quantitative model, we propose a novel 
bioinformatic method of data analysis for such modified experiment and apply 
it to extract the interaction parameters for a mammalian transcription factor  
CTF/NFI. From a practical point of view, our method results in a significantly 
improved false positive/false negative trade-off, as compared to both the 
standard information theory based method and a widely used empirically 
formulated procedure.
\end{abstract}

\bigskip

\hspace*{-0.73cm} \textbf{Keywords:} SELEX, protein-DNA interactions, 
transcription factor binding sites, weight matrix, CTF/NFI

\section{Introduction}
\label{S4.1}

One of the most important issues in molecular biology is to understand
regulatory mechanisms that control gene expression. Gene expression is often
regulated by proteins, called transcription factors (TFs), which bind to 
short (6 to 20 base pairs) segments of DNA~\cite{Lewin_2000}. To understand a 
regulatory system one needs a detailed knowledge of both TFs and their binding 
sites in a genome. Binding sites of a given TF share a common sequence 
pattern~\cite{Berg_1987}, which is often represented by a consensus sequence. 
However, TF binding sites are often highly degenerate, so it is not possible 
to reliably detect TF binding sites in a genome by using just the consensus 
sequence~\cite{Stormo_2000}. As an alternative, position-weight matrices 
(PWMs)~\cite{Berg_1987,Stormo_1982,Staden_1984} have been used to search for 
TF binding sites, with demonstrable advantage over consensus sequence based 
methods~\cite{Stormo_2000}.

The most widely used method to construct a PWM originates from
information-theoretic considerations~\cite{Staden_1984,Schneider_1986}. To 
distinguish such weight matrices from those constructed by other
methods, we will further call them information-theoretic weight matrices
(see also~\cite{Djordjevic_2003}). To build these weight matrices, one
usually starts from a known collection of aligned binding sites and
calculates the corresponding matrix elements as the logarithm of the ratio
of probability to observe a given base at a given position in a collection
of binding sites, compared to the probability of observing the base in the
genome as a whole~\cite{Stormo_2000}. However, despite the obvious advantages 
of using such PWMs over the consensus sequence, the majority of PWMs provide a 
low level of both sensitivity and specificity~\cite{Frech_1997}. In 
particular, there tends to be a large number of false positives in searches
using most PWMs~\cite{Stormo_2000,Frech_1997,Robinson_1998}.

In general, two problems may lead to the low sensitivity and specificity of
PWMs. First, the information-theoretic method may not be the most appropriate
one. It does not properly incorporate saturation in binding probability, as 
shown by~\cite{Djordjevic_2003}, and an alternative method of weight matrix 
construction\footnote{Weight matrices constructed by the method given 
in~\cite{Djordjevic_2003} were denoted energy matrices. This emphasizes that 
weights in the matrix correspond to the estimates of contribution to the 
binding (free) energy due to the presence of a certain base at a certain 
position in the binding site.}, based on the biophysical model of TF-DNA 
interaction, was developed. The method reduced the number of false positives 
and resulted in the explicit appearance - and determination - of the binding 
threshold. Additionally, there are probably problems with the 
collection of binding sites used to construct the weight (energy) 
matrix~\cite{Frech_1997} because, first, the collection of 
binding sites is most often obtained from a database, and is likely assembled 
under diverse and ill-characterized conditions~\cite{Djordjevic_2003}. Second, 
for most TFs, only a few binding sites are 
available~\cite{Wingender_2001,Salgado_2004}, making the amount of data 
insufficient for determining parameters of TF-DNA interaction (i.e. weight 
matrix).

As an alternative to using binding sites assembled in biological databases,
SELEX (Systematic Evolution of Ligands by Exponential Enrichment)
experiments~\cite{Tuerk_Gold_1990,Gold_1995} can be suitable for generating an 
appropriate dataset under controlled (uniform) conditions. Additionally, a 
recent experimental advance~\cite{Roulet_2002} which combines SELEX with SAGE 
(Serial Analysis of Gene Expression)~\cite{Valculescu_1995}, 
allows an efficient generation of a large number of binding sites for a given 
TF. In this paper, we ask the question: What is an appropriate experimental 
and computational procedure for inferring parameters of TF-DNA interaction 
from SELEX experiments? In particular, we will address the following two 
issues: 1) How should SELEX experiments be designed in order to 
generate a dataset suitable for determining parameters of TF-DNA interactions? 
2) How should a correct analysis of data from a suitable experiment be done? 
To address those questions, we will use a biophysical model of TF-DNA 
interactions to quantitatively model SELEX experiments. We will incorporate 
this model in the novel bioinformatic method of data analysis that we will 
subsequently develop.

The outline of this paper is as follows. In Section~\ref{S_WM}, we will review 
SELEX experiments and point to the potential problems in the experimental 
procedure from the viewpoint of constructing the appropriate weight matrix. In 
Section~\ref{S4.2} we will quantitatively model the SELEX experiments. Based 
on this model, we will show that there is a range of experimental parameters 
for which the energy matrix cannot be inferred by using the standard SELEX 
procedure. However, we will show that a modified experiment allows a robust 
generation of the appropriate data set. In Section~\ref{S4.3} we will propose 
a novel bioinformatic method of data analysis for modified SELEX experiments, 
and apply it to the data obtained in the experiment by Roulet 
{\em et al.}~\cite{Roulet_2002}. In Subsection~\ref{S4.3.2} we will show that 
our method leads to a significant improvement in the false positive/false
negative trade-off compared to the standard methods of data analysis. 
Quantitative analysis that supports Sections~\ref{S4.2} and~\ref{S4.3} is 
presented in Appendices B-F. Finally, in Section~\ref{S4.4} we will summarize
our results, compare them with some widely held views and put our work in the 
context of future research. 

\section{From SELEX to weight matrix}
\label{S_WM}

SELEX is a method in which a large number of oligonucleotides (DNA, RNA or 
unnatural compounds) can be rapidly screened for specific sequences that have
high binding affinities and specificities toward the given protein
target~\cite{Tuerk_Gold_1990}. For the explanation of the applications of 
SELEX procedure, one should refer to some of the review 
papers~\cite{Gold_1995,Gold_1997}. 

\begin{figure}[h]
\vspace*{6.4cm} \includegraphics{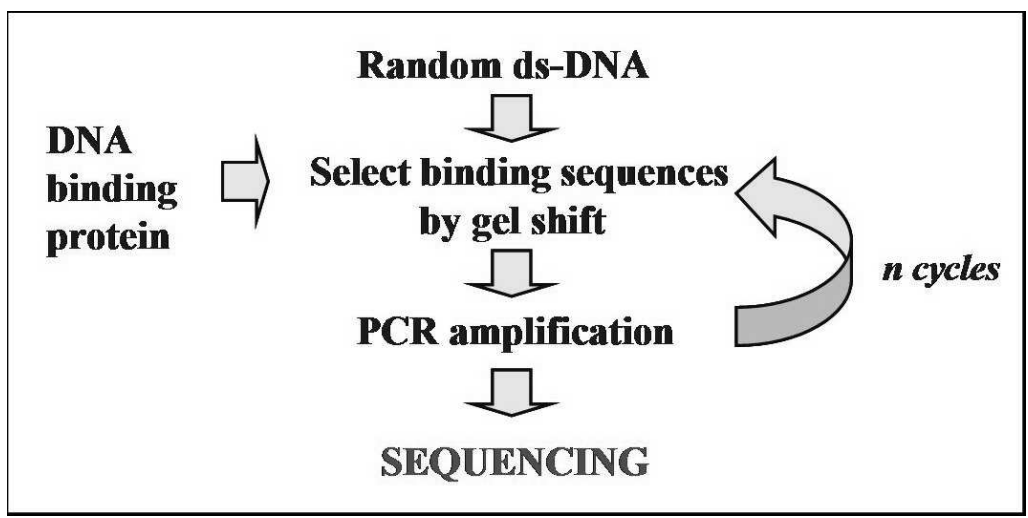} 
\caption{Scheme of SELEX experiment procedure. After n cycles of protein 
binding, selection and amplification, a certain number of DNA sequences are 
extracted and sequenced. Note that ds-DNA stands for double-stranded DNA.}
\label{SELEX_scheme}
\end{figure}

The scheme of the widely used SELEX experiment procedure is shown in 
Fig.~\ref{SELEX_scheme}. The experiment is usually performed as follows. In the
first step, a library of random oligonucleotides is synthesized. Protein is 
then mixed with the oligonucleotides library. Oligonucleotides that are bound 
by proteins are then separated from those that are not bound (e.g. by gel 
shift), which is called a selection step. Selected oligonucleotides are then 
amplified by the Polymerase Chain Reaction (PCR)~\cite{Saiki_1985}, which is 
called the amplification step. One cycle of TF binding, selection and 
amplification is called a SELEX round. The SELEX rounds are repeated several 
times~\cite{Tuerk_Gold_1990,Gold_1995}, and some number of sequences 
(typically from 20 to 50) are extracted and sequenced from the final round. 
This procedure is successful in identifying the strongest binding sites in the 
initial pool of random sequences, as demonstrated by experiments reported in 
the literature (e.g.~\cite{Tuerk_Gold_1990}), as well as by the numerical 
studies~\cite{Irvin_1991,VantHull_1998}. We will further call the widely used 
experimental procedure described above the {\it high stringency} SELEX, for 
reasons that will become apparent later. Next, in a typical data analysis 
the sequences selected in the last round of SELEX are used as a training set, 
from which elements of the information-theoretic weight matrix for a given TF 
are constructed~\cite{Stormo_1998,Cui_1995}. This weight matrix can then, in 
principle, be used to search for TF binding sites in a genome.  

Is the standard experimental and data analysis procedure outlined above really 
suitable to successfully infer a correct weight matrix? Looking at the
literature, it appears that this procedure fails often in practice. For 
example, in the SELEX experiment performed by Cui {\em et al.}~\cite{Cui_1995},
around 50 binding sites for LRP TF, selected in the last SELEX round, were 
extracted and sequenced. Binding sites were then used to construct an 
information-theoretic weight matrix, and the binding dissociation constants of 
the extracted sequences were then experimentally measured. However, the 
correlation between the dissociation constants and the information scores 
(i.e. the weight matrix scores) was quite poor, and accordingly, Cui 
{\em et al.}~\cite{Cui_1995} comment that ``the poor correlation for the data 
described here seems surprising...''. Further, in~\cite{Robinson_1998} a 
comprehensive comparison between the weight matrices obtained from the eight 
available SELEX experiments with {\em E. coli} TFs and the corresponding 
weight matrices constructed from natural binding sites was performed. In seven 
out of those eight cases, large discrepancies between the matrices derived 
from natural sites and those derived from SELEX were reported. Therefore, 
obtaining good weight matrices from the standard SELEX procedure appears to be 
more an exception than a rule.

Why does the procedure described above appear to fail in many cases? The first
possibility is that the assumption of additivity in TF-DNA 
interactions~\cite{Benos_2002}, on which the weight matrix representation is 
based, may not be suitable. This is, however, not likely, since this 
approximation has proved to be very good in many 
cases~\cite{Stormo_1998,Benos_2002,Takeda_1989,Sarai_1989}. Alternatively, the 
reason for the failure might be that the analysis was done by using 
the information theory based method. As discussed in the Introduction, 
the information theory based procedure does not properly incorporate 
saturation in binding probability. Therefore, it is not surprising that this 
is not the most appropriate method for correctly inferring the energy matrix. 
In this paper, we will develop a method of data analysis based on the 
biophysical model of protein-DNA interactions. 

However, it seems rather surprising that the information theory based method 
is the only reason for the apparent problems with the weight matrices 
inferred from SELEX discussed above. There may be a 
systematic problem with the high stringency SELEX procedure when it comes to 
generating a data set suitable for inferring weight/energy matrices. 
With regard to this, it is apparent that two possible problems may arise. 
First, it may be that the noise in the dataset is too large, i.e. that many
of the extracted sequences are too weak or are non-specific binders 
(for a discussion on nonspecific binding see Appendix B). Second, if the
extracted sequences consist of only the strongest binding sites, the inferred 
weight matrix elements will come with large errors.  To observe this, it is 
useful 
to take the limit in which only the sequence corresponding to the consensus 
binding site is extracted, where it is obvious that the energy matrix cannot 
be obtained from such information. For a more detailed statistical analysis of 
this issue refer to~\cite{Roulet_2002}. Therefore, our first
goal is to address possible systematic problems with the experimental 
procedure by quantitatively modeling the SELEX experiments. We also 
incorporate this model in the bioinformatic method of data analysis that 
we develop in Section~\ref{S4.3}.

\section{A quantitative model of SELEX}

\label{S4.2}

Our model of SELEX is based on the biophysical view of TF-DNA interactions, 
which was used in a few recent papers (see 
e.g.~\cite{Djordjevic_2003,Gerland_2002,Sengupta_2002}). 
For completeness and for introducing the notation, we briefly review a 
biophysical model of TF-DNA interaction in Appendix A. We start this section 
by extending this model in several ways, as to make it suitable for modeling 
of SELEX.

First, we take into account non-specific binding of a TF to DNA. As we show
in Appendix B,  the binding probability (Eq.~(\ref{Eq_bind_prob})) is 
(approximately) modified in the following way due to the non-specific binding 
(see Eqs.~(\ref{Eq_bind_pr}) and~(\ref{Eq_b_prob})): 
\beqar
p\left( S\right) \approx \frac{1}{\exp \left( E\left( S\right) -\mu \right)+1}
+c_{ns}=f\left( E\left( S\right) -\mu \right) +c_{ns},
\eeqar{Eq_ns_binding}
where $E(S)$ is binding (free) energy\footnote{For brevity, from now on we 
will refer to the free energy of binding simply as ``binding energy''. In 
chemical literature, the commonly used notation for this quantity would be 
$\Delta G(S)$ rather than $E(S)$.} of TF to a DNA sequence $S$, $\mu$ is 
chemical potential (see Appendix A), while $c_{ns}$ depends on the threshold 
of non-specific binding $E_{ns}$ (see Eq.~(\ref{Eq_gamma})). Note that we 
scale all energies with $k_B T$. From Eq.~(\ref{Eq_ns_binding}) follows that 
non-specific binding cannot be distinguished from the so-called background 
partitioning.
Background partitioning~\cite{Tuerk_Gold_1990} is an effect that, during the
selection step (Section~\ref{S4.1}), it is not possible to perfectly
separate sequences that are bound by protein from those that are not bound.
Due to that, in each round of SELEX, some DNA sequences {\it not} bound by
TF are also selected with probability $c_{b}$. A combined effect of
non-specific binding and background partitioning can be described by 
Eq.~(\ref{Eq_ns_binding}), where $c_{ns}\rightarrow c_{ns}+c_{b}$ (further in 
the text, we denote $c=c_{ns}+c_{b}$). Although $c$ itself is likely small 
(e.g. $c_{b} \sim 0.1 \%$ in~\cite{Tuerk_Gold_1990}), the effect of 
non-specific binding/background partitioning can be considerable, since in 
SELEX experiments DNA is typically in large excess over protein, so only a 
small fraction of all DNA sequences are typically (specifically) bound by TF.  

Second, we take into account that the length of DNA sequences in SELEX
experiments is usually larger than the length of the TF binding site. For 
example, in the experiment by Roulet {\em et al.}~\cite{Roulet_2002}, which 
will be the subject of our analysis in Section~\ref{S4.3}, the sequence length 
is $l=25$ bp long, while the length of the binding site for CTF/NFI TF, 
studied in the experiment, is $L=15$ bp. In Appendix C we discuss the 
modification of the binding probability due to the fact that $l$ is greater 
than $L$. We show that this effect can be approximately accounted for by 
modifying the distribution of binding energies from $\rho(E)$ (see 
Eq.~(\ref{Eq_Rho_E})) to $\rho _{M}\left( E\right) $ given by 
Eq.~(\ref{Eq_Rho_M}). Additionally, we note that the model has to take into 
account that the support $E_{S}$ of the energy distribution is finite (see 
Eq.~(\ref{Eq_support})), with ``bottom of the band'' determined by the energy 
of the strongest binder in the pool of random DNA sequences.

We note that we neglect stochastic effects in our model. This is
generally justified by the fact that the typical length $L$ of TF binding
site is 20bp or less, while the total number of sequences used in SELEX is 
typically $N\sim 10^{15}$~\cite{Gold_1995}, so each possible DNA sequence
of length $L$ is present in about $N/4^{L}\sim 10^{3}$ copies. Additionally, 
we note that high fidelity RNA polymerase is used in SELEX experiments, so 
mutations of sequences during the PCR amplification can be generally 
neglected. In that respect, SELEX is different from the so-called 
{\em in vitro} evolution experiments~\cite{Libchaber,Peng_2003}, where low 
fidelity RNA polymerase and larger number of PCR rounds are used to 
(purposefully) introduce mutations and generate strong binding sequences that 
do not exist in the initial, small ($N \sim 10^5$), random DNA pool. 

\subsection{High stringency SELEX}

\label{S4.2.1}

In this subsection, we model the high stringency SELEX procedure, based on the 
(extended) model of TF-DNA interactions described above. We introduce 
equations that allow us to
determine the position of chemical potential $\mu ^{\left( n\right) }$ and
energy distribution of selected oligos $\rho _{M}^{\left( n\right) }\left(
E\right) $, as a function of the number of performed SELEX rounds $n$: 
\beqar
\rho _{M}^{\left( n\right) }\left( E\right) \sim \left[ f\left( E-\mu
^{\left( n\right) }\right) +c \right] \,\rho _{M}^{\left( n-1\right)
}\left( E\right)  \label{Eq_Rho_n}
\eeqar{Eq_S4.3}
and 
\beqar
p_{t}&=&p_{f}^{\left( n\right) }+p_{b}^{\left( n\right) }=K\exp \left( \mu
^{\left( n\right) }\right) +  \nonumber \\
&&+\int \rho _{M}^{\left( n-1\right) }\left( E\right)
\,f\left( E-\mu ^{\left( n\right) }\right) \,dE+d_{t}\,\exp( \mu^{(n)}-E_{ns})
\label{Eq_mu_n}
\eeqar{Eq_S4.4}

In the equations above, $p_{t}$, $p_{f}^{\left( n\right) }$ and $%
p_{b}^{\left( n\right) }$ are respectively total, free, and bound
concentrations of protein and $d_{t}$ is the total amount of DNA. 
Eq.~(\ref{Eq_Rho_n}) connects energy
distributions of selected sequences from n$^{th}$ and (n-1)$^{th}$ round.
Eq.~(\ref{Eq_mu_n}) is mass conservation law, and it determines the
position of $\mu ^{\left( n\right) }$. $\rho _{M}^{\left( n-1\right) }\left(
E\right) $ in Eq.~(\ref{Eq_mu_n}) is normalized to $d_{t} $. Note that all 
energies are rescaled by $k_B T$. The equations
above are solved recursively, i.e. we first solve them for $n=1$, then
increase $n$ by one etc. Note that $\rho _{M}^{\left( 0\right) }\left(
E\right) $ is $\rho _{M}\left( E\right) $ given by Eq.~(\ref{Eq_Rho_M}) 
(Appendix C). We also note that, since in SELEX experiments DNA is typically
in (large) excess over protein ($d_{t}\gg p_{t}$), most of the protein is 
bound to DNA and $p_{f}^{\left( n\right) }$ (i.e. 
$K\exp \left( \mu ^{\left( n\right)}\right) $ term in Eq.~(\ref{Eq_mu_n})) 
can be neglected compared to $p_{b}^{\left( n\right) }$.

In general, Eqs.~(\ref{Eq_Rho_n}) and~(\ref{Eq_mu_n}) have to be solved 
numerically; however the main features can be understood qualitatively. Let us 
assume that total amount of protein and the total amount of DNA (after each 
amplification step) are kept constant in each round of the experiment (as e.g. 
in~\cite{Tuerk_Gold_1990}). It is evident that as $n$ increases, the 
(average) affinity of selected oligos will also increase, which leads to the 
increase in the amount of $p_{b}^{\left( n\right) }$ and decrease of 
$p_{f}^{\left( n\right)}$. Therefore, both $\mu ^{\left( n\right) }$ (note 
that $\mu^{(n)}=\log(p_f^{(n)}/K)$) and the maximum $E_{m}^{\left( n\right) }$ 
of energy distribution $\rho _{M}^{\left( n\right)}\left( E\right) $ move to 
stronger binding energies with the increase in the number of  performed SELEX 
rounds. (Many experiments are performed in a way that the total amount of 
protein $p_{t}\equiv p_{t}^{\left( n\right) }$ is decreased from one SELEX 
round to the next (e.g. see~\cite{He_1996}). It is evident that the previous 
conclusion about the decrease of $\mu ^{\left(n\right) }$ holds in this case 
as well.) A limit in which Eqs.~(\ref{Eq_S4.3}) and~(\ref{Eq_S4.4}) can be 
solved analytically is analyzed in Appendix D. In particular, 
Eqs.~(\ref{Eq_max_n_pos}) and~(\ref{Eq_mu_j}) quantitatively support the 
discussion above. 

\begin{figure}
\vspace*{6.4cm} \includegraphics{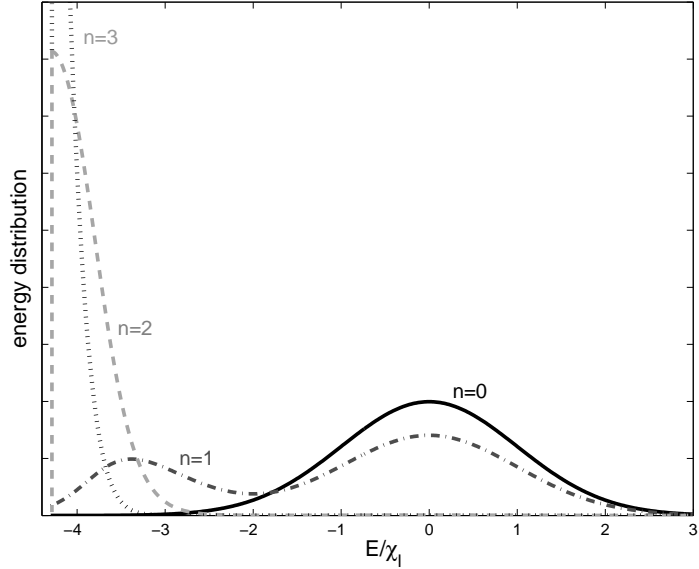} 
\caption{Change of $\rho _{M}^{\left( n\right) }\left( E\right) $ for a 
different number of performed rounds $n$, for a high stringency SELEX 
experiment. Peaks centered around zero correspond to random DNA binders, while 
the left hand corner of the figure corresponds to the highest affinity binder.}
\label{high_strin_SELEX}
\end{figure}

In further analysis, we take the following values of parameters: $\chi
_{l_{e}}=4\,k_{B}T$ (see Appendix C), $p_{t}=10nM$, $d_{t}=10\mu M$ and 
$E_{S}=-4.3\chi_{l_{e}} $, while $E_{ns}=-2.0\chi_{l_{e}} $ (see Appendix B). 
The assumed values of $p_{t} $ and $d_{t}$ are typical for SELEX 
experiments~\cite{Tuerk_Gold_1990}, while (typical) $L \in \{6,...,20\}$ bp 
(base pairs) leads to $E_S$ in the interval from $-3.5\chi_{l_{e}}$ to 
$-7\chi_{l_{e}}$ (see Eq.~(\ref{Eq_support})). 
Values of $\chi_{l_{e}}$ and $E_{ns}$ are expected to differ from one TF to 
the other, but the assumed values are likely inside the realistic 
range~\cite{Gerland_2002}. Figure~\ref{high_strin_SELEX} shows 
$\rho_{M}^{\left( n\right)}\left(E\right)$ numerically obtained from 
Eqs.~(\ref{Eq_Rho_n}) and~(\ref{Eq_mu_n}), with the parameter values 
stated above. It is useful to observe how the signal to noise ratio changes
with $n$, where the noise is the number of selected random binders 
(corresponding to peak centered around zero), while the signal is the number 
of selected specific binders (for signal to noise ratio in the limit of 
unsaturated binding see Appendix D and Eq.~(\ref{Eq_signal_noise})). For 
$n=1$, we see that there is a small number of specific binders (note the small 
peak centered at $E/\chi_{l_{e}}\approx -3.5$) compared to the number of 
selected random sequences, so that signal to noise ratio is low. Because of 
the high noise to signal ratio, it is not possible to infer (correct) energy 
matrix from such a dataset. For $n=2$, random binders are completely 
eliminated, so the problem with noise does not exist anymore. However, another 
problem emerges, i.e. since energy distribution of selected oligos has reached 
support $E_{S}$, (only) the strongest binding sites are selected. As 
discussed in Section~\ref{S_WM}, a correct energy matrix cannot be obtained 
from such a sequence set. For $n=3$, we select the sequence set with even 
stronger binding affinities, etc. We note that the sharp cuts in distributions 
$\rho^{(n)}(E)$ at $E_{S}$ (see Fig.~\ref{high_strin_SELEX}) are the 
consequence of the approximation that we use for $\rho_{M}^{(0)}(E)$ (see 
Eq.~(\ref{Eq_cut1})). In reality, $\rho_M^{(n)}(E)$ becomes discrete when one 
approaches $E_S$, which obviously does not change the conclusions inferred 
from Fig.~\ref{high_strin_SELEX}.

We solved Eqs.~(\ref{Eq_S4.3}) and~(\ref{Eq_S4.4}) for different parameter 
values. As shown by the example above, for a range of realistic parameter 
values the appropriate choice for the total number of performed rounds $n$ 
does not exist\footnote{In the analysis presented here, we used that $p_t$ is 
constant in each round of SELEX. If $p_t$ is decreased from one round to the 
next, as it is, in practice, done in many experiments (e.g.~\cite{He_1996}), 
it is evident that this problem is even more pronounced.}. On the other hand, 
for some (other) parameters, for which selected DNA sequences have an 
acceptable signal to noise ratio and $E_{m}^{(n)}$ does not reach the highest 
affinity binders, the optimal choice of $n$ does exist. However, we note that 
in practice, it is very hard to reliably predict such $n$, i.e. to decide when 
to stop the experiment, because a studied TF has {\em a priori} unknown 
binding parameters (i.e. $\chi _{l_e}$ and $E_{ns}$). Therefore, even in the 
case when the appropriate choice of $n$ exists, it cannot, in practice, be 
simply calculated from Eqs.~(\ref{Eq_Rho_n}) and~(\ref{Eq_mu_n}). From this, 
it follows that the high stringency SELEX procedure is, in practice, unsuitable
for inferring the parameters of TF-DNA interaction. In the next section, we 
discuss modification of the SELEX procedure that allows a robust generation of 
the appropriate set of binding sequences.

\subsection{SELEX with fixed selection stringency}
\label{S4.2.2}

Let us now assume that instead of moving toward the stronger binding energies,
the chemical potential $\mu ^{(n)}=\mu $ is constant in each round of the 
experiment. In such a case, from Eq.~(\ref{Eq_Rho_n}) it follows that the 
energy distribution is given by the following (simple) expression: 
\begin{eqnarray}
\rho _{M}^{\left( n\right) }\left( E\right) \sim f^{n}\left( E-\mu \right)
\,\rho _{M}^{\left( 0\right) }\left( E\right)  \label{Eq_dist_unsat}
\end{eqnarray}

\begin{figure}[tbp]
\vspace*{6.4cm} \includegraphics{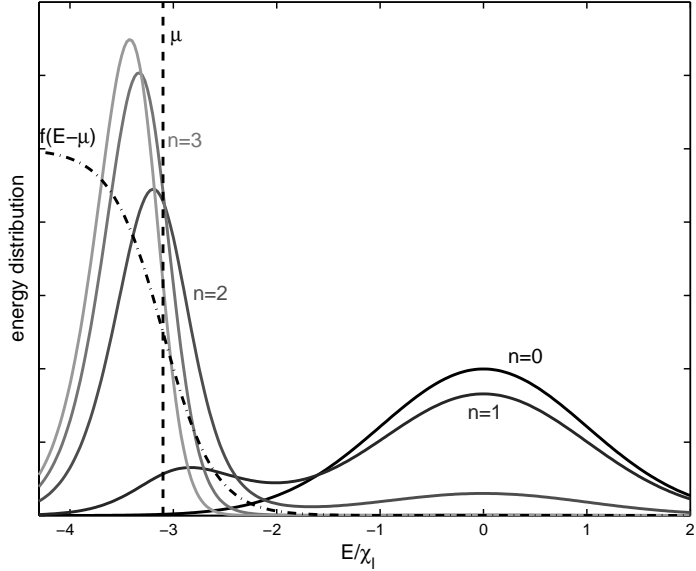} 
\caption{Change of $\rho _{M}^{\left( n\right) }\left( E\right) $ for different
number of performed rounds $n$, for a SELEX experiment in which the chemical 
potential $\mu$ is fixed. Note that once the maximum of 
$\rho _{M}^{\left( n\right)}\left( E\right) $ reaches $\mu $, it further moves 
very slowly toward the higher binding affinities. $\mu = -3.1\chi _{l_e}$, 
while the other parameters are the same as in Fig.~\ref{high_strin_SELEX}.}
\label{fixed_strin_SELEX}
\end{figure}

As we show in Appendix E, from Eq.~(\ref{Eq_dist_unsat}) it follows that 
$E_{m}^{\left( n\right) }=\overline{E}_{l_{e}}-n\chi _{l_{e}}^{2}$ in the 
first rounds of experiment, when $n<(\overline{E}_{l_{e}}-\mu) 
/\left( \chi _{l_{e}} \right) ^{2}$. Thus, in the first few rounds, the 
maximum of energy distribution $E_{m}^{\left( n\right) }$ rapidly moves to the 
higher affinities. However, once $E_{m}^{\left( n\right) }$ reaches $\mu $, it 
further drifts very slowly toward stronger binding energies 
(Eq.~(\ref{Eq_Max_corr})). Figure~\ref{fixed_strin_SELEX} is equivalent 
to Fig.~\ref{high_strin_SELEX} with the difference that 
$\mu ^{\left( n\right)}=\mu$ is fixed throughout the experiment. We
see that for $n=1$, we have the same situation as in the high stringency
experiment, i.e. the signal to noise ratio is too low. For $n=2$, non-specific
binders are eliminated, similarly as in the high stringency experiment.
However, the important difference is that instead of reaching the strongest
binders (i.e. $E_{S}$), $E_{m}^{\left( 2\right) }$ is close to $\mu $. For $%
n>2$, $E_{m}^{\left( n\right) }$ drifts very slowly toward the higher
binding energies, remaining in the proximity of $\mu $, and consequently does 
not reach $E_{S}$. More precisely, in Appendix E we show that $E_m^{(n)}$
asymptotically approaches $\mu - 2 k_B T$.

The procedure described above has a significant practical advantage compared
to the high stringency experiment discussed in the previous subsection. Since 
$E_{m}^{\left( n\right) }$ remains essentially fixed for larger $n$, one can
perform more rounds (say 4 or 5), thus being sure that random binders are 
eliminated, without the risk that only the strongest sequences will be 
selected. Since the procedure tolerates the whole range of $n$ (in the 
above example $n>2$), we call it robust.

The next issue is how the constraint of fixed chemical potential can be 
experimentally implemented. To our knowledge, all but one of the performed 
experiments correspond to high stringency SELEX. However, in the experiment 
done by Roulet {\em et al.}~\cite{Roulet_2002}, the SELEX experiment was 
modified by inclusion of the radiolabeled sequence (probe) of moderate binding 
affinity $E^{\ast }$. The concentration of the DNA, added to the reaction 
mixture as a competitor to the radiolabeled probe, was in each round adjusted, 
so that a fixed fraction of the probe is bound by CTF/NFI TF in each SELEX 
round. From this it follows that 
$f\left( E^{\ast }-\mu ^{\left( n\right)}\right)$
$=const$, leading to $\mu ^{\left( n\right) }=const$. Therefore the procedure 
in~\cite{Roulet_2002}, provides a practical solution for fixing chemical 
potential through different rounds of experiment.

We also note that the analysis above gives a practical criterion at what $n$ 
the experiment should stop. The procedure can be completed when random binders 
are eliminated and $E_m^{(n)}$ has reached $\mu$, at which point we have a 
``quasi-saturation'' (see Fig.~\ref{fixed_strin_SELEX} and 
Eq.~(\ref{Eq_Max_corr})). Since at quasi-saturation the total amount of DNA 
(adjusted and directly observed by experimentalists) ceases to significantly 
change from one round to the next, this gives a practical criterion at what 
$n$ the experiment should end. 

\section{SELEX data analysis}

\label{S4.3}

In this section we present a bioinformatic method for the data analysis of
fixed stringency ($\mu=const$) SELEX experiments.  We will first briefly 
present the basic idea behind our method. We will then introduce our novel 
algorithm for constructing energy matrix in Subsection~\ref{S4.3.1}. In 
Subsection~\ref{S4.3.2} we will apply the algorithm to the experimental data, 
and compare the results with both the information theory based method and a 
widely used empirically formulated procedure, MatInspector~\cite{MatInspector}. 

Figure~\ref{fixed_strin_SELEX} and Appendix E show that the maximum of energy 
distribution for oligos selected in the final rounds of SELEX has to be in the 
vicinity of the chemical potential. It follows that the majority of sequences 
extracted from SELEX are in the saturated regime, i.e. bound with the 
probability close to one (see Appendix A). In~\cite{Djordjevic_2003} we showed 
that the information theory method is appropriate to use when sequences 
are in the unsaturated regime, but that this method does not properly 
incorporate saturation in the binding probability. In the context of SELEX 
experiments considered here, the information theory based method would be 
appropriate to use only if the majority of oligos were in the exponential 
tail of the binding probability $f(E-\mu)$. Since this does not happen in the 
fixed stringency SELEX experiments, we will in this Section devise a method 
which uses correct binding probability. Additionally, as described in 
Section~\ref{S4.2} and Appendix C, we have extended a biophysical model of 
TF-DNA interactions to take into account that the length of TF binding site 
is typically shorter than the lengths of DNA sequences in SELEX, which we will 
also incorporate into our new procedure.     

A key point in the implementation of our method will be to obtain TF-DNA 
interaction parameters through a maximum likelihood procedure. We will infer 
the initially unknown parameters by maximizing the probability that the 
extracted set of DNA sequences is observed as the outcome of the experiment. 
The probability of extracting the given set of DNA sequences will be 
calculated by taking into account the correct TF-DNA binding probability 
(which properly describes the saturation effects) and by appropriately 
modifying affinity distribution of DNA oligos to account for difference in 
lengths between TF binding sites and SELEX DNA sequences. The set of equations 
resulting from varying this probability with respect to the unknown parameters 
will be then numerically solved to compute the elements of the energy matrix.

\subsection{Estimating the energy matrix}

\label{S4.3.1}

In this subsection, we introduce a novel algorithm appropriate for data 
analysis of fixed stringency SELEX experiments. Let us assume that after $n$ 
rounds of SELEX, set $A$, which contains $n_{S}$
sequences $S^{\left( j\right) }$ ($j\in \left( 1,..,n_{S}\right) $), has
been extracted and sequenced. As summarized above, we will infer the unknown 
energy matrix by the maximum likelihood procedure. The probability of observing
sequences from set $A$, but no other sequences from the initial DNA pool, is 
given by: 
\beqar
\exp \left( \Lambda \right) = \prod_{S\in A}\gamma \,f^{n}\left( E\left(
S\right) -\mu \right) \,\prod_{S^\prime \notin A}[1-\gamma \,f^{n}\left(
E\left( S^\prime \right) -\mu \right) ]  \label{Eq_L}
\eeqar{Eq_S4.6}
Terms $f^{n}$ in the equation above account for $n$ selection processes. The 
factor $\gamma $ is a ``sampling probability'' and it absorbs
all extraction and amplification events (there are $n$ of them), as well as 
the final sequencing after the $n^{th}$ round. Probabilities of extraction, 
amplification and sequencing are all assumed to be independent of sequence 
$S$, so $\gamma $ does not depend on $S$ either. We also note 
that we have neglected non-specific binding/background partitioning in 
Eq.~(\ref{Eq_L}), since we assume that $n$ in Eq.~(\ref{Eq_L}) is large 
enough, so that non-specific binders are eliminated by the $n^{th}$ round (see 
Fig.~\ref{fixed_strin_SELEX}). Further, the sum over unbound sequences 
$S^{\prime }$ can be approximated in terms of the binding energy distribution: 
\beqar
\prod_{S^\prime \notin A}[1-\gamma \,f^{n}\left( E\left( S^{\prime }\right)
-\mu \right) ]&\approx &\exp [-\gamma \,\sum_{S^\prime \notin O}f^{n}\left(
E\left( S^\prime \right) -\mu \right) ]  \nonumber \\
&& \hspace*{-3.cm} \approx \exp [-\gamma \int \rho _{M}\left( E\right)
\,f^{n}\left( E-\mu \right) \,dE  \label{Eq_Rho_1}]
\eeqar{Eq._S4.7}

In the above equation, we use $\rho _{M}\left( E\right) $ (see 
Eq.~(\ref{Eq_Rho_M})) instead of $\rho \left( E\right) $ (see 
Eq.~(\ref{Eq_Rho_E})) to account for the fact that the length of used DNA 
sequences is typically larger than the length of the TF binding site (see 
Appendix C). Similarly, we approximate $f(E(S)-\mu)$ in the first term of 
Eq.~(\ref{Eq_S4.6}) by $f(E(s_M)-\mu)$ (see Appendix C), where $s_{M}$ is the 
TF binding site of length $L$, with the maximal binding energy on $l$ long 
sequence $S$. In practice, the set of binding sites $s_M$ can be identified by 
an unsupervised search of the set of sequences $S$ for $L$ long statistically 
overrepresented motifs (e.g. by using the Gibbs search 
algorithm~\cite{Lawrence_1993}). We also note that in the first (approximate) 
equality in Eq.~(\ref{Eq._S4.7}), we used $\gamma \ll 1$, which is justified by
the fact that the number of DNA sequences with binding energies below $\mu$ is
typically much larger than the number of sequences $n_S$ extracted from the 
last round of the experiment. For example, if we assume $\mu=-3 \chi_{l_e}$ 
(as in Fig.~\ref{fixed_strin_SELEX}), $n_S \sim 10^3$~\cite{Roulet_2002} and 
typical $N \sim 10^{15}$~\cite{Gold_1995}, we have 
$\gamma =n_S/[\int f^n (E-\mu)\rho_M (E) dE] \sim 10^{-9}$.

Since changing the overall scale of energy corresponds to multiplying energy 
scores for all binding sites with the (same) constant, for bioinformatic 
purposes, i.e. for TF binding site identification, it is not necessary to 
determine the overall scale of energy (see also~\cite{Djordjevic_2003}). The 
natural quantity to scale all energies is width $\chi$ (in units of $k_B T$)
of energy distribution for an ensemble of random oligos of length $L$ 
(see~\cite{Djordjevic_2003} as well as 
Eqs.~(\ref{Eq_Rho_E}-\ref{Eq_chi_square}) in Appendix A). With such scaling, 
and provided that zero of energy is set to coincide with the mean 
$\overline{E}$ of energy distribution in the ensemble of random sequences (see 
Eq.~(\ref{Eq_mean_E}) in Appendix A), $E(S)/\chi$ directly gives the estimate 
of significance of the given energy score. That is, the probability that a 
random DNA sequence will have a stronger binding energy than $E(S)/\chi$ is 
given by 
\beqar
\int_{-\infty}^{E(S)/\chi} \rho(E) dE \approx \frac{1}{2}
[1-erf(-\frac{E(S)/\chi} {\sqrt{2}})] \sim exp(-E(S)^2/(2\chi^2))
\eeqar{Eq._erf} 
Here $\rho(E)$ is the energy distribution for the set of random oligos 
(Eq.~(\ref{Eq_Rho_E}) in Appendix A), $erf(x)$ is the error function, and the 
last approximation is valid for $|E(S)/\chi| \gg 1$. Also note that if we 
consider the energy matrix as an vector in $4 L$ dimensional space, $\chi$ is 
equal to the norm of this vector (see Eq.~(\ref{Eq_chi_square1}) in Appendix 
A), so rescaling with $\chi$ corresponds to normalizing the energy matrix to 
unit ``length''. We further use the notation 
$\widetilde{\epsilon }_{i,\alpha }=\epsilon _{i,\alpha }/\chi $. 

Additionally, since maximum $E_m^{(n)}$ (i.e. the mean within the gaussian 
approximation of $\rho _{M}^{\left( n\right) }\left( E\right) $) has to
be close to $\mu $ (see Subsection~\ref{S4.2.2} and 
Fig.~\ref{fixed_strin_SELEX}), we impose the constraint that 
\begin{eqnarray}
\mu /\chi =\sum_{S\in A}E\left( S\right) /\left( n_{S}\,\chi \right)
=\sum_{i,\alpha }\widetilde{\epsilon }_{i,\alpha }\,S_{i,\alpha }^{\ast }
\label{Eq_constr}
\end{eqnarray}
where $S_{i,\alpha }^{\ast }=\sum_{j}S_{i,\alpha }^{\left( j\right) }/n_{S}$. 
In order to obtain $\widetilde{\epsilon }_{i,\alpha }$ we maximize $\Lambda$
(defined by Eqs.~(\ref{Eq_S4.6}),~(\ref{Eq._S4.7}) and~(\ref{Eq_constr})) with 
respect to $\widetilde{\epsilon }_{i,\alpha }$ and $\gamma $. Variation of 
$\Lambda $ with respect to $\widetilde{\epsilon }_{i,\alpha }$ and $\gamma$ 
then leads to the set of equations which are given in Appendix F. Those 
equations can be numerically solved to obtain 
$\widetilde{\epsilon }_{i,\alpha }$.   

\bigskip 
\subsection{Application of the algorithm}
\label{S4.3.2}

To demonstrate our method, we use it to analyze the data from the experiment 
by Roulet {\em et al.}~\cite{Roulet_2002}. In addition to the modification 
discussed in Subsection~\ref{S4.2.2}, Roulet {\em et al.} combined SELEX with 
the SAGE protocol~\cite{Valculescu_1995}, 
which allowed them to sequence a large number of DNA oligos. A large dataset 
provides an obvious advantage for a precise estimation of energy parameters. 
In particular, a total of 4 SELEX rounds were performed, and 
approximately 880, 960, 1200, 6900 and 230 sequences were obtained from 
rounds 0, 1, 2, 3, and 4 respectively (note that round 0 refers to the 
initial, completely random, DNA pool).

We use about 230 sequences obtained after the $4^{th}$ round of experiment, to
estimate $\widetilde{\epsilon }_{i,\alpha }$. We first search those 230 
sequences $S$, to identify CTF/NFI binding sites $s_M$ that correspond to
$L=15$ bp long statistically overrepresented motifs. The unsupervised search 
was performed by using a Gibbs search based algorithm~\cite{Bioprospector}, 
and the obtained set of binding sites $s_M$ was used to determine 
$\widetilde{\epsilon}_{i,\alpha}$ by numerically solving Eq.~(\ref{Eq_final}). 
The obtained energy matrix $\widetilde{\epsilon}_{i,\alpha}$ is given in 
Appendix F (Table 1). We will further call $\widetilde{\epsilon}_{i,\alpha}$ 
the finite $T$ energy matrix, to emphasize that, contrary to the QPMEME 
algorithm~\cite{Djordjevic_2003}, the optimization of $\Lambda$ 
(Eq.~(\ref{Eq_S4.6})) is not done in $T \rightarrow 0$ limit. We shift columns 
of $\widetilde{\epsilon }_{i,\alpha }$\footnote{Note that a 
provisional base independent constant can be added to each column of weight 
(energy) matrix, which corresponds to shifting zero of the weight matrix 
scores.}, so that the mean of the energy distribution $\rho \left(E\right)$ 
(given by Eq.~(\ref{Eq_mean_E})), for the ensemble of random oligos of length 
$L$ is zero.

\vspace{2cm}

\begin{figure}
\vspace*{8.9cm} \includegraphics{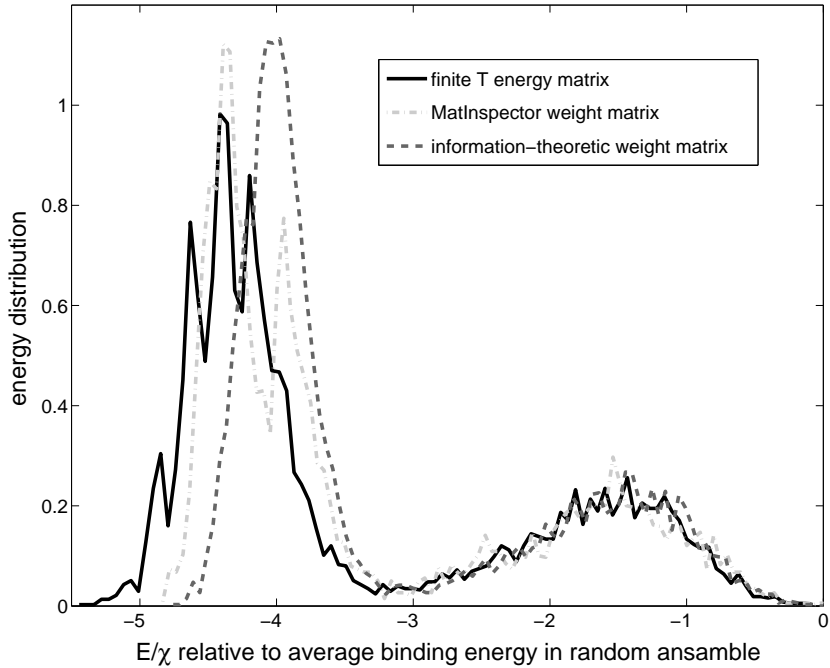} 
\caption{ Energy distributions obtained by using the finite T energy
matrix, the information-theoretic weight matrix and the MatInspector weight
matrix are compared. Energy
distributions are computed for more than 6000 binding sites extracted from the
$3^{rd}$ round of SELEX~\cite{Roulet_2002}. Zero on the horizontal axis, 
corresponds to the mean value of binding energy in random ensemble. Note that 
maximum of energy distribution for non-specific binders is displaced relative 
to zero (i.e. positioned at around -1.5), caused by the fact that the length 
of the DNA sequences (25bp) is larger than the length of CTF/NFI TF binding 
site (15 bp). Actually, the energy distribution of non-specific binders, 
matches well with $\rho _{M}\left( E\right) $ calculated in Appendix F 
(Eq.~(\ref{Eq_Rho_M})).}
\label{energy_distribution}
\end{figure}

We next compare our method with both the information theory based method and a 
widely used empirically formulated procedure, MatInspector~\cite{MatInspector}.
For this purpose, we construct both the information theory weight matrix 
$w_{i,\alpha}$ (see e.g.~\cite{Hertz_1999}) and the MatInspector weight matrix 
$w_{i,\alpha}^{MI}$ (see~\cite{MatInspector}) from the same set of binding 
sites $s_M$ that we used to compute our finite T energy matrix. The obtained 
matrices are given and compared in Appendix F. We normalize $w_{i,\alpha }$ and
$w_{i,\alpha}^{MI}$ and choose the zeros of weight matrix scores (i.e. 
``energies'') for both matrices in the same way as for 
$\widetilde{\epsilon }_{i,\alpha}$. The three matrices are then used to 
compute the corresponding energy distributions for more than 6000 sequences 
extracted from the $3^{rd}$ round of SELEX, and their comparison is shown in 
Fig.~\ref{energy_distribution}. We see that there is a noticeable difference 
in the estimates of energies obtained by the three weight matrices.

\begin{figure}[tbp]
\vspace*{8.4cm} \includegraphics{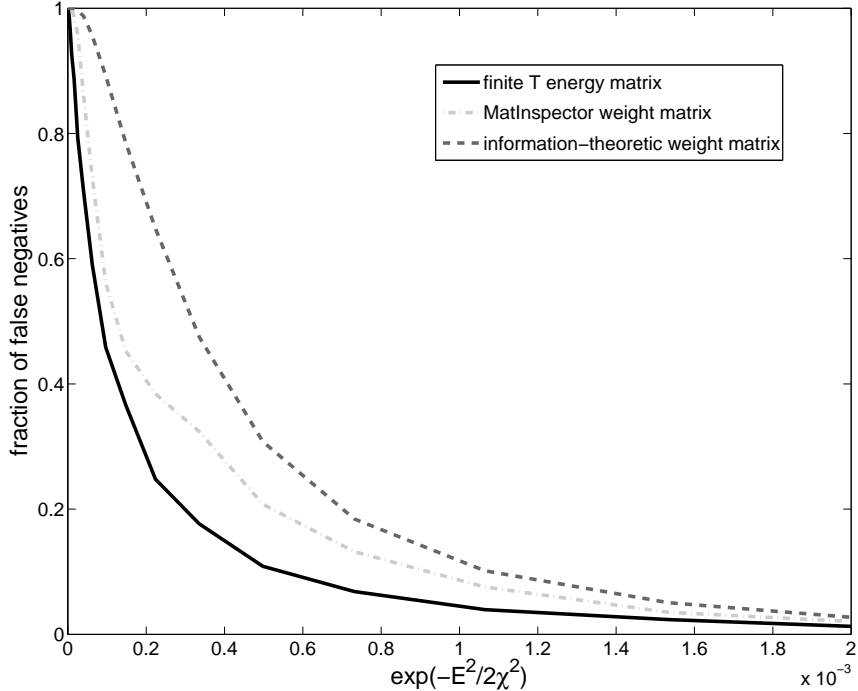} 
\caption{Comparison of the DET curves for the finite T energy matrix, the 
information theory weight matrix and the MatInspector weight matrix. Note that 
$E/\chi$ is measured relative to the average binding energy in random ensemble,
so that $\exp(-E^2/(2\chi^2))$ on the horizontal axis is proportional to the 
number of false positives. False negative fraction is inferred from 
Fig.~\ref{energy_distribution}, by calculating cumulatives of the 
corresponding energy distributions.}
\label{DET_curve}
\end{figure}

To compare performance of the methods, we will infer false positive/false 
negative trade-off for the three matrices. The fraction of false negatives, 
for certain threshold $E/\chi$, can be readily estimated by computing 
cumulatives of the distributions shown in Fig.~\ref{energy_distribution}. 
Further, the fraction of false positives can be calculated by computing the 
corresponding cumulative of the energy distribution of random oligos (see 
Eq.~(\ref{Eq_Rho_E}) in Appendix A). Therefore, with our choice for the zero 
of energy, it is evident that $\exp(-E^2/(2\chi^2))$ (on the abscissa of 
Fig.~\ref{DET_curve}) is proportional to the number of false 
positives. More precisely, if one searches a random DNA sequence with 
total length $N_{D}$, the total number of binding sites with energy scores 
below $E/\chi $ is approximately $(N_{D}/\sqrt{2\pi }) 
\int_{-\infty }^{E/\chi}\exp ( -x^2/2) d x \sim N_D \exp(-E^2/(2\chi^2))$ 
(see Eq.~(\ref{Eq_Rho_E}) and Eq.~(\ref{Eq._erf})), where the last 
approximation is valid for $|E/\chi| \gg 1$. 

The above estimates of false positives and false negatives can be used to 
obtain the Detection Error Trade-off (DET) curves 
(see~\cite{Djordjevic_2003,Martin_1997}), which are shown in 
Fig.~\ref{DET_curve}\footnote{In the construction 
of the DET curve for MatInspector we did not use the so-called ``core 
similarity", which can optionally be used as a second threshold in the 
method~\cite{MatInspector}. The reasons are that the use of core similarity 
is left as an ``optional feature at the discretion of the 
user"~\cite{MatInspector}, which was not used in the example searches 
in~\cite{MatInspector}. Additionally, in~\cite{MatInspector} it is not stated 
how to choose the value of the second threshold, which makes it hard to fairly 
compare a method with two (arbitrary) thresholds with the methods that use 
only one threshold.}. We see that over the entire range of fixed false 
negative values on the vertical axis, fraction of false positives for the 
finite T energy matrix is few times lower compared to the false positive rate 
for the information-theory weight matrix. In the case of the MatInspector 
weight matrix, the finite T energy matrix has about two times smaller false 
positive rate for false negative fractions smaller than $40\%$, while for 
larger false negative fractions the two DET curves come close to each other. 
We note that the meaningful range, in which one is likely to operate in 
practice, is the one with a smaller fraction of false negatives (i.e. $40\%$ 
or below), where the finite T energy matrix clearly shows better performance.

Figure~\ref{DET_curve} shows that, while our method outperforms both 
MatInspector and the information theory method, it appears that MatInspector 
shows a better false positive/false negative trade-off compared to the 
information theory method. We however note that MatInspector is an empirically 
formulated procedure, i.e. it is not founded on either statistical 
considerations, such as the information theory method, or on a (bio)physical 
model of TF-DNA interactions, such as the method presented here. We therefore 
believe that the performance of MatInspector would change from one experiment 
to the other, and that a better performance of MatInspector compared to the 
information theory method may not prove to be systematic. On the other hand, 
since our method is based on a correct physical model of SELEX experiments, we 
expect that it will systematically produce reliable estimates of the 
interaction parameters. In any case, fixed stringency SELEX experiments, which 
will be performed in future, will provide opportunities to more throughly test 
the performance of the algorithms.       

Finally, we discuss how the above comparison of the three 
algorithms may be affected by the possible presence of noise in the input data 
used to construct the three weight matrices. We obtained the set of (putative) 
binding sites through a heuristic Gibbs sampling algorithm, which is in 
principle not guaranteed to find the true binding sites. So if there are 
misassignments in the input data, the question is whether the better 
performance of 
our method could be the result of higher robustness with regard to noise, 
rather than higher accuracy for error-free data. To asses the amount of noise 
in the input data we performed a self-consistency check. We scored all DNA 
sequences selected in the $4^{th}$ round of the experiment with the finite T 
energy matrix and classified all binding sites above score $-3.4$ as specific 
binders and all below $-3.4$ as non-specific binders (note that $-3.4$ 
corresponds to the saddle of the bimodal distribution shown in 
Fig.~\ref{energy_distribution}). We find that only one in the 175 large set of 
input binding sites obtained by the Gibbs algorithm is not contained in the 
list of 184 specific binders classified with the finite T energy matrix. This 
indicates that the noise level in the input data is likely quite small. The 
apparent low number of misassignments in the output of Gibbs alignment is 
likely a consequence of the low number of random sequences extracted from 
$4^{th}$ round of SELEX\footnote{Notice that Fig.~\ref{energy_distribution} 
shows that there is a quite small amount of noise even in the $3^{rd}$ round 
of experiment. The noise is further reduced in the $4^{th}$ round.}, and the 
fact that the length of SELEX sequences is quite short ($25$ bp).

\section{Conclusion and Outlook}
\label{S4.4}

In this paper we modeled SELEX experiments and proposed a novel method
of data analysis. Our analysis showed that for a certain realistic range of 
parameters, the suitable solution for the number of rounds that should be 
performed does not exist at all. We argued that even for the parameters for 
which the solution exists, it is very hard to find such a solution in practice.
However, we showed that the modification of the standard SELEX procedure in 
which chemical potential is fixed~\cite{Roulet_2002} robustly selects 
sequences that allow one to successfully determine the energy matrix. We next 
proposed a novel method for inferring energy matrix from the sequences 
extracted from a SELEX experiment with fixed selection stringency. Contrary 
to the widely used information theory weight matrix method, our procedure 
correctly represents saturation in binding probability. As an example of the 
procedure, we estimated the energy matrix for CTF/NFI TF by analyzing the 
data from the experiment by Roulet {\em et al.}~\cite{Roulet_2002}. We 
demonstrated that our energy matrix leads to a significantly better false 
positive/false negative trade-off.

Finally, we compared the results of our analysis with some widely held
views. It is generally well understood that doing too large a number of SELEX
rounds leads to too strong a selection. It is, however, widely believed that
the problem can be solved by doing only few selection cycles. For example,
in the recent SELEX experiment by Kim {\em et al.}~\cite{Kim_2003}, it was 
noted that ``...we used fewer rounds of selection than a conventional SELEX to
avoid the enrichment of just a few high affinity winners''. However, as
shown by our analysis, this can lead to another problem, i.e. if too few 
SELEX rounds are performed, too large a number of random sequences are likely 
to be selected. Moreover, we found that the problems of over-selection and too 
high amount of noise are, in practice, very hard to reconcile within the 
standard SELEX procedure and that a modified experiment with a fixed chemical 
potential has to be performed instead. From the aspect of data analysis, we 
showed that the commonly used information theory based method, widely believed 
to be well founded in both statistics and thermodynamics~\cite{Fickett_2000}, 
is not appropriate for analysis of data for SELEX experiments with fixed 
selection stringency. 

In the context of future research, we believe that the analysis presented 
here, together with the experimental methods introduced in~\cite{Roulet_2002}, 
open a perspective to apply high-throughput, fixed stringency, SELEX 
experiments for a large number of different transcription factors. This would 
provide a reliable method for detection of transcription factor binding sites, 
and would facilitate the comprehensive understanding of gene regulation.

\section*{Acknowledgments} 

This work was supported by NIH grant GM67794 (to Boris Shraiman). Final parts 
of this work were supported by NSF under Agreement No. 0112050 and NSF grant 
MCB-0418891. We are grateful to Boris Shraiman for useful discussions, and to 
Andrew Millis, Harmen Bussemaker and Istok Mendas for critical reading of the 
manuscript. M.D. acknowledges the hospitality of the Kavli Institute for 
Theoretical Physics where part of this work was done and is grateful to 
Valerie Parsegian for reading the manuscript.

\section*{Glossary}

\hspace*{0.6 cm}{\em Weight matrix \hspace{0.2 cm}} for a binding pattern of 
length L is defined as a matrix of numbers $w_{i,\alpha}$ where 
$i \in [1,2,...L]$ and $\alpha \in [A,T,C,G]$. The score of the sequence 
$\alpha_1 \dots \alpha_L$ is given by $w_{1,\alpha_1} + w_{2,\alpha_2} + 
\cdots + w_{L,\alpha_L}$.

\medskip

{\em Sensitivity \hspace{0.2 cm}} is defined as TP/(TP+FN), where TP is the number of true 
positives and FN is the number of false negatives. In the context of a TF 
binding site search, a true positive (TP) arises when an algorithm correctly 
classifies a true binding site as such, while a false negative (FN) arises 
when an algorithm classifies a true binding site as a non-binding site.

\medskip
                                                                    
{\em Specificity \hspace{0.2 cm}} is defined as TN/(TN+FP), where TN is the number of true 
negatives and FP is the number of false positives. In the context of TF 
binding site search, a true negative arises when an algorithm correctly 
classifies a true non-binding site as such, while a false positive (FP) arises 
when a search algorithm classifies a true non-binding site as a binding site.

\medskip

{\em Serial Analysis of Gene Expression (SAGE)\hspace{0.2 cm}} is a method for 
comprehensive analysis of gene expression patterns. In the context of this 
paper, a part of the SAGE protocol can be used to link together oligomers 
extracted from SELEX in order to form longer DNA molecules that can be 
efficiently sequenced.  

\medskip

{\em Gel shift\hspace{0.2 cm}} is a technique used to separate free DNA 
molecules from DNA molecules that are in complex with protein, based on the 
fact that protein-DNA complexes migrate more slowly through gel under the 
influence of an electric field. 

\medskip

{\em Polymerase Chain Reaction (PCR)\hspace{0.2 cm}} is an experimental 
technique that allows one to produce a large number of copies of any fragment 
of DNA. In principle, the number of DNA molecules is doubled in each round of 
PCR, so there is an exponential increase in the number of molecules with the 
number of performed PCR rounds. 

\medskip

{\em Dissociation constant\hspace{0.2 cm}} for sequence $S$ is equal to the 
concentration of (free) TF for which there is $50\%$ probability that a DNA 
molecule $S$ will be bound by the TF. The relationship between the 
dissociation constant and the binding energy $E(S)$ is given by 
$K_D(S) = K \exp(E(S))$, where $K$ is a proportionality constant.

\medskip

{\em Maximum likelihood estimation\hspace{0.2 cm}} is a statistical method 
used to estimate unknown parameters of a (known) probability distribution. The 
basic principle is to draw a sample from the distribution, calculate the
probability that this sample is observed and then determine the unknown 
parameters such that this probability is maximal.

\begin{appendix}

\section{A biophysical model of TF-DNA interaction}

Let us consider an experiment where a certain number of identical DNA
oligomers with sequence $S$ and length $L$ (equal to the length of the TF
binding site) are mixed into a solution with some concentration of TF. It
can be shown (see e.g.~\cite{Djordjevic_2003}) that the equilibrium 
probability $p(S)$ that a DNA sequence $S$ is bound by TF is given by:

\beqar
p(S)=\frac{1}{1+\exp ( E( S)-\mu ) }=f( E( S) -\mu )
\eeqar{Eq_bind_prob}
In the above equation, $p_{f}$ is the concentration of free TF, $\mu $ is
the chemical potential, while $K$ is a multiplicative constant related with 
counting number of quantum states in a box (e.g. see~\cite{Kubo_book}).

Chemical potential $\mu $ is set by the free TF concentration in the
solution:

\beqar
\mu =\log ( p_{f}/K) .
\eeqar{Eq_chem_pot}
Note that in the above equations all energies are rescaled by $k_{B}T$. The
form of the binding probability $f( E( S) -\mu ) $ in
Eq.~(\ref{Eq_bind_prob}) corresponds to the Fermy-Dirac distribution (see 
e.g.~\cite{Kubo_book}). If the binding energy $E(S)$ of a sequence $S$ is well 
below $\mu $, then $f(E( S) -\mu )$ is close to one and the 
sequence $S$ is almost always bound by TF. We will further call sequences with 
binding energy $E(S)$ which corresponds to this limit saturated. On the other 
hand, if $E(S)$ is well above $\mu $, the sequence $S$ is rarely bound, with 
probability given by the Boltzmann distribution $f(E(S)-\mu) \approx 
\exp (-(E(S)-\mu))$. As shown in~\cite{Djordjevic_2003}, the 
information theory weight matrix procedure assumes Boltzmann distribution of 
binding probability, and is, therefore, not appropriate whenever saturation 
of binding occurs. 

Further, we need an expression for $E(S)$. The most simple model of TF-DNA 
interaction, which we use in this paper, assumes that the interaction of a 
given base with the factor does not depend on the neighboring bases:

\beqar
E( S) \approx \epsilon \cdot S=\sum_{i=1}^{L}\sum_{\alpha
=1}^{4}\epsilon _{i}^{\alpha }S_{i}^{\alpha },
\eeqar{Eq_bind_energy}
where $S_{i}^{\alpha }=1$, if base $\alpha $ is at the position $i$ and 
$S_{i}^{\alpha }=0$ otherwise. $\epsilon _{i}^{\alpha }$ is the interaction
energy with the nucleotide $\alpha $ at the position $i=1,...,L$ of the DNA
string~\cite{Stormo_1998}, and $\epsilon $ is called energy matrix. The simple
parameterization given by Eq.~(\ref{Eq_bind_energy}) provides a very good 
approximation in many 
cases~\cite{Stormo_1998,Takeda_1989,Benos_2002,Sarai_1989}, although there are
examples where binding at some positions in the binding site shows
dependence on dinucleotide pairs~\cite{Man_and_Stormo_2001,Bulyk_2002,O'Flanagan_2005}.

We further need to compute the energy distribution $\rho (E)$
for an ensemble of randomly generated oligonucleotides. Such ensemble
corresponds to the random set of DNA sequences used in the first round of
SELEX experiment. In~\cite{Djordjevic_2003} it was shown that in the first 
approximation $\rho (E)$ is given by a gaussian:

\beqar
\rho ( E) \approx \exp ( -(E-\overline{E})^{2}/2\chi
^{2}) /\sqrt{2\pi \,\chi ^{2}}
\eeqar{Eq_Rho_E}
with

\beqar
\overline{E}=\sum_{i=1}^{L}\overline{\epsilon }_{i}
\eeqar{Eq_mean_E}
and
\beqar
\chi ^{2}=\sum_{i=1}^{L}\sum_{\alpha =1}^{4}p_{\alpha }
(\epsilon_{i}^{\alpha }-\overline{\epsilon }_{i})^{2}
\eeqar{Eq_chi_square}
where $\overline{\epsilon }_{i}=\sum_{\alpha =1}^{4}p_{\alpha}
(\epsilon_{i}^{\alpha }) $.

As noted in footnote 4, each column of an energy matrix can be shifted for a 
provisional base independent value, and a convenient choice would be to set 
$\overline{\epsilon }_{i} = 0$, so that $\overline{E} = 0$ and 
\beqar
\chi ^{2}=\sum_{i=1}^{L}\sum_{\alpha =1}^{4}p_{\alpha }
(\epsilon_{i}^{\alpha })^{2}\; .
\eeqar{Eq_chi_square1}

From the above equation follows that $\chi$ is equal to the norm of the
energy matrix $\epsilon_{i}^{\alpha }$, with ``metric'' given by the 
background single base frequencies  $p_{\alpha }$.

We further note that $\rho ( E) $ is well approximated by Eq.~(\ref{Eq_Rho_E})
in the proximity of maximum $E=\overline{E}$, however, away from the maximum
deviations from gaussianity appear. In fact, the support of $\rho (E)$ is
finite with the ``bottom of the band'' $E_{S}=\sum_{i}\min_{\alpha}\,
\epsilon _{i,\alpha }$ (while the ``top'' is given by $\sum_{i}\max_{\alpha
}\epsilon _{i,\alpha }$). In this paper we work with the Gaussian
approximation to $\rho ( E) $, but we also introduce a cut in the distribution 
to account for the fact that the support $E_{bb}$ is finite (see Appendix C).

Further extensions of the TF-DNA interaction model, necessary for our
modeling of SELEX experiments, are given in Appendix B and Appendix C.

\section{Non-specific binding of TF to DNA}

We here assume that a given TF can bind to DNA in two conformations. The
first conformation results in the sequence specific interaction, with the
interaction energy $E\left( S\right) $. The second conformation results in
the sequence independent (non-specific) interaction, with the interaction
energy $E_{ns}$~\cite{Berg_1987}. $E_{ns}$ is called the threshold
for non-specific binding. We consider a reversible reaction of binding of
the TF to a DNA sequence $S$, where TF can bind with $S$ in two
conformations. Sequence specific, and sequence non-specific reactions can,
respectively, be represented by:
\beqar
\left[ TF\right] +\left[ S\right] \Leftrightarrow \left[ TF-S\right] _{s}
\eeqar{Eq_re_s}

and
\beqar
\left[ TF\right] +\left[ S\right] \Leftrightarrow \left[ TF-S\right] _{n}
\eeqar{Eq_re_n}

Here, $\left[ TF\right] $ is concentration of free TF, $\left[ S\right] $ is
concentration of sequence $S$ that is \textit{not} in the complex with
protein, while $\left[ TF-S\right] _{s}$ and $\left[ TF-S\right] _{n}$ are
concentrations of TF that is bound to the TF in the sequence specific, and
in the sequence non-specific conformation respectively. In the equilibrium,
the following relations hold:
\beqar
K \exp \left( E\left( S\right) \right) =\frac{\left[ TF\right] \,
\left[ S \right] }{\left[ TF-S\right] _{s}}
\eeqar{Eq_equ_S}
\beqar
K \exp \left( E_{ns}\right) =\frac{\left[ TF\right] \,\left[ S\right] }{%
\left[ TF-S\right] _{n}}
\eeqar{Eq_equ_n}

From Eqs.~(\ref{Eq_equ_S}) and~(\ref{Eq_equ_n}) we have that the probability 
that a sequence $S$ is bound by the TF is given by:
\beqar
p\left( S\right) =\frac{\left[ TF-S\right] _{s}+\left[ TF-S\right] _{n}}{%
\left[ S\right] +\left[ TF-S\right] _{s}+\left[ TF-S\right] _{n}}=\frac{a}{%
b\exp \left( E\left( S\right) \right) +1}+c_{ns}.
\eeqar{Eq_bind_pr}

In the equation above,
\beqar
a=\frac{1}{1+\exp \left( \mu -E_{ns}\right) }
\eeqar{Eq_a}
\beqar
b=\exp \left( -\mu \right) \,\left[ 1+\exp \left( \mu -E_{ns}\right) \right]
\eeqar{Eq_b}

and
\beqar
c_{ns} =\frac{1}{1+\exp \left( E_{ns}-\mu \right) }
\eeqar{Eq_gamma}
where $\mu =\log \left( \left[ TF\right] /K\right) $. By comparing 
Eq.~(\ref{Eq_bind_pr}), with Eq.~(\ref{Eq_bind_prob}) the quantity $b$ can be 
identified as the ''effective'' fugacity in the presence of non-specific 
binding. From Eq.~(\ref{Eq_b}) then follows that, non-specific binding, in 
principle, shifts $\mu $ toward more negative values. In practice, however, we 
most often have a case in which the amount of TF is much less than the amount 
of DNA. For example, even for a pleiotropic TF such as CRP in {\em E. coli}, 
the total number of CRP molecules is much less than the total length of the 
genome~\cite{Anderson_1971}, while in SELEX experiments protein is typically 
in large excess over DNA. It is then obvious that $\mu $ has to be 
significantly below $E_{ns}$, since all DNA sequences would, otherwise, have 
to be bound with high probability. Therefore, we in practice have that 
$\exp \left( \mu -E_{ns}\right) \ll 1$, so from Eqs.~(\ref{Eq_a}) 
and~(\ref{Eq_b}) follows that $a\approx 1$, $b\approx \exp \left(-\mu \right)$
and $c_{ns} \approx \exp \left( -\left( E_{ns}-\mu \right)\right) $, 
so we have that:
\beqar
p\left( S\right) \approx \frac{1}{\exp \left( E\left( S\right) -\mu \right) 
+1}+c_{ns} 
\eeqar{Eq_b_prob}

Therefore, the effect of non-specific binding enters through $c_{ns} $, which 
is determined by the position of $\mu $ relative to $E_{ns}$. We note that 
values of $E_{ns}$ were not experimentally quantitated~\cite{Gerland_2002}, 
so one does not know what range of values $E_{ns}$ can take. Unless $E_{ns}$ is
positioned in the strong binding tail of $\rho \left( E\right) $, non-specific 
binding will not have a large effect.  As discussed in Section~\ref{S4.2}, 
non-specific binding is in SELEX experiments essentially indistinguishable 
from the background partitioning effect.

\section{Binding of TF to a longer DNA sequence}

In Appendix A, we derived the binding probability under the assumption that 
the length $l$ of DNA sequence is equal to the length $L$ of the TF binding
site. However, for DNA sequences used in SELEX, $l$ is typically larger than 
$L$. For example, in the SELEX experiment performed by Roulet 
{\em et al.}~\cite{Roulet_2002} (see Subsection~\ref{S4.2.2}) $l=25$ bp, while 
$L=15$ bp for CTF/NFI TF. It is straightforward to obtain that the probability 
that sequence with length $l$ will be bound is given by:
\beqar
p(S) = \frac{\exp (\mu) \sum_{i=1}^{l-L}\exp (-E(s_{i}))}{1+\exp(\mu) \,
\sum_{i=1}^{l-L}\exp(-E ( s_{i}))} 
\eeqar{Eq_bprob}

In the equation above, the sequences $s_{i}$ are $L$ long binding sites,
corresponding to all possible $l-L$ frame shifts in which the TF can bind to 
a sequence $S$, while $\mu $ is chemical potential (see Appendix A). 
We assume that $l<2\,L$, which is typically the case in SELEX experiments, so
that two or more TF molecules cannot simultaneously bind to the sequence $S$. 
Note that all quantities in Eq.~(\ref{Eq_bprob}) are rescaled by $k_{B}T$.
For $l_{e}=l-L$ that is not too large, which is typically the case in the
experiments, the expression $\sum_{i=1}^{l_{e}}\exp \left( -E\left(
s_{i}\right) \right) $ can be approximated by taking into account only the
contribution from the strongest binding site $s_{M}$, where $E_{M}\left(
S\right) =E\left( s_{M}\right) =\min \{E\left( s_{i}\right) ,$ $i\in \left(
1,l_{e}\right) \}$. With this approximation, Eq.~(\ref{Eq_bprob}) simplifies
to:
\beqar
p(S) \approx f( E_{M} (S)-\mu) =\frac{1}{1+\exp(E_{M} (S)-\mu)} \,
\eeqar{Eq_bprob_app}

which is the Fermi Dirac probability encountered before (see Appendix A).

From Eq.~(\ref{Eq_bprob_app}) follows that binding of a TF to a sequence $S$ 
with $l>L$ is (approximately) equivalent to the binding to the sequence $s_{M}$
with length $L$. Let us now look at the first round of SELEX, where the TF is 
mixed with a large number of randomly generated sequences of length $l$. In 
order to make the problem equivalent to the one in which $l=L$, instead of 
density of states $\rho \left( E\right) $ (see Appendix A), we have to use 
$\rho _{M}\left( E\right)$ defined as the number of sequences $S$, for which 
$E_{M} (S)$ has energy between $E$ and $E+dE$.

To calculate $\rho _{M}\left( E\right) $, we neglect correlations in binding
energies $E\left( s_{i}\right) $ (see Eq.~(\ref{Eq_bprob})) of $l_{e}$
binding sites that belong to the same sequence $S$. This is in general well
justified, unless sequence $S$ consists of the long repeat. In particular, the 
validity of this approximation was confirmed by numerically testing 
Eq.~(\ref{Eq_Rho_M}) below. Based on this, $\rho _{M}\left( E\right) $
can be calculated by generating sets of $l_{e}$ values of $E$ from
distribution $\rho \left( E\right)$ and retaining only the strongest
binding energy from each set. It is straightforward to see that 
$\rho_{M}(E^{\prime})$ can be obtained from $\rho(E)$ by:
\beqar
\rho _{M}\left( E^{\prime}\right) =l\,_{e}\frac{d\Phi \left( E^{\prime}\right) 
}{dE^{\prime}}\,\left(1-\Phi \left( E^{\prime}\right) \right) ^{l_{e}-1},
\eeqar{Eq_cut}
where $\Phi \left( E^{\prime}\right) $ is the cumulative distribution given by 
$\Phi\left( E^{\prime}\right) =\int_{-\infty }^{E^{\prime}}\rho( E) \,dE$. The 
expression on the right hand side of Eq.~(\ref{Eq_cut}) is,
therefore, equal to the probability that $l_{e}-1$ values of $E$ generated
from $\rho (E)$ are above $E^{\prime }$, while one of them (the strongest
binding energy) is between $E^{\prime }$ and $E^{\prime }-dE^{\prime }$.

By plotting $\rho _{M}\left( E\right) $, given by Eq.~(\ref{Eq_cut}) we see 
that $\rho_{M}(E)$ can be approximated by Gaussian:
\beqar
\rho _{M}\left( E\right) \approx \exp \left( \left( E-\overline{E}_{l_{e}}
\right) ^{2}/2\chi _{l_{e}}^{2}\,\right) ,
\eeqar{Eq_Rho_M}
with 
\beqar
\overline{E}_{l_{e}}=\overline{E}-a\left( l_{e}\right) \,\chi 
\eeqar{Eq_M_mean} 
and 
\beqar
\chi_{l_{e}}=b\left( l_{e}\right) \,\chi ,
\eeqar{Eq_M_chi}
where $\overline{E}$ and $\chi $ are respectively the mean value and the
standard deviation for $\rho \left( E\right) $ (see Appendix A),
while $a\left( l_{e}\right) $ and $b\left( l_{e}\right) $ are respectively
monotonically increasing and decreasing functions of $l_{e}$. Functions $%
a\left( l_{e}\right) $ and $b\left( l_{e}\right) $ can be calculated
numerically (e.g. $a\left( 10\right) =1.4$ and $b\left( 10\right) =0.3$).
Numerical analysis shows that approximately, $a\left(
l_{e}\right) \approx 0.6\,\log \left( l_{e}\right) $, while $b\left(
l_{e}\right) \approx 1/\sqrt{l_{e}}$.

Finally, we have to take into account that $\rho _{M}\left( E\right) $ has
the finite support, where ''bottom of the band'' $E_{S}$ is determined by the 
energy of the strongest binder in the random pool of DNA sequences, and can be 
approximated by:
\beqar
(4^{L})\int_{-\infty }^{E_{S}} \rho \left( E\right) \,dE\,\sim \,1,
\eeqar{Eq_support}
where $\rho \left( E\right)$ is normalized to 1.

In Eq.~(\ref{Eq_support}), we assumed that the total number of sequences $N$
(more precisely $l_{e}\,N$) in the DNA pool is larger than $4^{L}$, so that all
possible sequences of length $L$ are present. This is most often the case in 
practice, since typically $L<20$, while $N \sim 10^{15}$~\cite{Gold_1995}, so 
that $4^{L} \ll N$.

To take into account the finite support of $\rho _{M}\left( E\right) $, we 
make a simple approximation and introduce a sharp cut in the distribution 
$\rho _{M}\left(E\right) $, i.e. we take that
\beqar
\rho _{M}\left( E\right) \sim \theta\left( E-E_{s}\right) \exp \left(
\left( E-\overline{E}_{l_{e}}\right) ^{2}/2\chi _{l_{e}}^{2}\,\right) ,
\eeqar{Eq_cut1}
where $\theta\left( E-E_{s}\right) $ is unit step (Heaviside) function. We 
note that the top of the band is finite as well, however we do not include it 
in Eq.~(\ref{Eq_cut1}), since energy distribution of selected oligos moves 
toward higher binding affinities in SELEX (see Subsection~\ref{S4.2.1}). In 
reality, $\rho_{M}\left( E\right)$ becomes discrete when we approach $E_{s}$, 
however a simple approximation given by Eq.~(\ref{Eq_cut1}) is sufficient for 
the purpose of our model.

\section{High stringency SELEX in the limit of unsaturated binding}

In this Appendix, we look at the limit in which the binding probability 
$f\left( E-\mu \right) $ in Eqs.~(\ref{Eq_S4.3}) and~(\ref{Eq_S4.4}) can be 
approximated by the Boltzmann factor $\exp (\mu -E)$. In this limit, 
those equations can be solved analytically. The above 
approximation is valid if all selected binding sites are unsaturated in each 
SELEX round, i.e. if $\mu ^{(k)}<E_{S}$ $(\forall k\in (1,..,n))$, where $n$ 
is the number of performed SELEX rounds. In this limit, Eq.~(\ref{Eq_S4.3})
gives:
\beqar
\rho _{M}^{(n)}(E)\sim(\exp (-E)+\exp (-E_{ns}))^{n}\,\rho _{M}(E),
\eeqar{Eq_Rho_n_unsat}
where, for simplicity of the notation, we assume that all noise comes from
non-specific binding, i.e. that $c\equiv c_{ns}$ (see 
Eqs.~(\ref{Eq_ns_binding})). If we use the gaussian approximation for 
$\rho_{M}(E)$ (see Eq.~(\ref{Eq_Rho_M})), from Eq.~(\ref{Eq_Rho_n_unsat}) 
follows that $\rho _{M}^{(n)}(E)$ has $n$ peaks, which are centered at 
positions $E_{m,k}^{(n)}=-k\chi_{l_{e}}^{2}$, $(k\in (1,..n))$. We here 
shifted zero of energy, so that it coincides with $\overline{E}_{l_{e}}$ 
(see Eq.~(\ref{Eq_M_mean})). From Eq.~(\ref{Eq_Rho_n_unsat}) it is obvious 
that $n^{th}$ peak $E_{m,k=n}^{(n)}\equiv E_{m}^{(n)}$ contains only sequence 
specific binding sites, while $k=0$ peak (centered at zero) corresponds 
exclusively to non-specific (i.e. random) binders. Therefore, in the limit 
considered here, the maximum of specifically selected binding sites (the 
leading maximum) is positioned at:
\beqar
E_{m}^{(n)}=-n\,\chi _{l_{e}}^{2}.
\eeqar{Eq_max_n_pos}

From Eq.~(\ref{Eq_max_n_pos}) follows that $E_{m}^{(n)}$ rapidly moves
to higher binding energies. For example, for the realistic parameter values
of $\chi _{l_{e}}=2$ and $E_{S}=-5\chi _{l_{e}}$ (corresponding to $L=12$, 
see Eq.~(\ref{Eq_support})), $E_{m}^{(n)}$ reaches $E_{S}$ after (only) three 
SELEX rounds. 

The ''intensity'' $I_{k}^{(n)}$ of $k^{th}$ peak, i.e. the number of binding
sites corresponding to the peak, is:
\beqar
I_{k}^{(n)} &\sim& \exp (-E_{ns}(n-k))\,\int \exp \left( -Ek\right)
\rho _{M}(E)\, \nonumber \\
&\approx& \exp (-E_{ns}(n-k))\exp (k^{2}\chi _{l_{e}}^{2}/2)
\eeqar{Eq_int_k}

From Eq.~(\ref{Eq_int_k}) is straightforward to obtain that the conditions 
$I_{k}^{(n)}>I_{n}^{(n)}$ and $I_{k}^{(n)}>I_{0}^{(n)}$ cannot be
simultaneously satisfied for $k\in \left( 2,...,n-1\right) $, for any
parameter values. Therefore, for any given $n$, either $0^{th}$ or $n^{th}$
peak have the maximal intensity. Further, it is sensible to define ''signal
to noise'' ratio $\nu ^{(n)}$ as the ratio of the number of specific binders
corresponding to the $n^{th}$ peak and the number of non-specific binders
corresponding to the $0^{th}$ peak. From Eq.~(\ref{Eq_int_k}) follows:
\beqar
\nu ^{(n)}=\exp (n^{2}\chi _{l_{e}}^{2}/2+n\,E_{ns}),
\eeqar{Eq_signal_noise}
so signal to noise ratio necessarily increases with increasing $n$. 

We next want to derive for what parameter values is the unsaturated limit,
which we analyze in this Appendix, valid. We use the self-consistency 
condition: 
\beqar
\mu ^{(j)}<E_{S} \, (\forall j\in (1,...n)),
\eeqar{Eq_nonsat_cond}
with $\mu ^{(j)}$ calculated from Eq.~(\ref{Eq_S4.4}) by using  
$f(E-\mu)\approx \exp (\mu -E)$:
\beqar
\exp (\mu ^{(j)})=\frac{p_{t}/d_{t}}{\exp [(j-1/2)\chi
_{l_{e}}^{2}\,]\,\alpha (j,\chi _{l_{e}},E_{ns})+\exp (-E_{ns})}
\eeqar{Eq_mu_j}
where,
\beqar
\alpha (j,\chi _{l_{e}},E_{ns})=\frac{1+\sum_{k=0}^{j-2}{j-1 \choose k}
[\exp(-(k+j+1)\chi _{l_{e}}^{2}/2-E_{ns}]^{j-k-1}}{1+\sum_{k=0}^{j-2}
{j-1 \choose k}[\exp (-(k+j-1)\chi _{l_{e}}^{2}/2-E_{ns}]^{j-k-1}}.
\eeqar{Eq_alpha_1}

\medskip

We note that the condition (\ref{Eq_nonsat_cond}) (with $\mu^{(j)}$ given by 
Eqs.~(\ref{Eq_mu_j}) and~(\ref{Eq_alpha_1})) is satisfied for the
subset of parameter values that are inside the realistic range. For example,
the condition (\ref{Eq_nonsat_cond}) holds for $p_{t}=10\,nM$, $%
d_{t}=10\,\mu M$ (see e.g.~\cite{Tuerk_Gold_1990}), $\chi _{l_{e}}=2\,
$, $E_{ns}=-2\chi _{l_{e}}$, $E_{S}=-5\chi _{l_{e}}$, and for all $n$ until $%
E_{m}^{(n)}$ reaches $E_{S}$ (i.e. for $n=1,2,3$).

\medskip

We finally discuss how $\mu ^{(n)}$ depends on the parameter values (for
further discussion, we let $j\rightarrow n$ in Eq.~(\ref{Eq_mu_j})). From
Eq.~(\ref{Eq_mu_j}) is straightforward to show that $\mu ^{(n)}$ necessarily
decreases, with increasing $n$. Further, the following interpretation
can be assigned to the terms on the right hand side of Eq.~(\ref{Eq_mu_j}).
It is obvious that $\mu ^{(n)}$ increases, with the increase of (total)
protein to (total) DNA ratio $p_{t}/d_{t}$. The term $\exp [(n-1/2)\chi
_{l_{e}}^{2}\,]$ accounts for the fact that the (average) binding energy of
sequences corresponding to the $n^{th}$ peak increases with $n$ (see 
Eq.~(\ref{Eq_max_n_pos})), which results in the decrease of the concentration 
of \textit{free} TF and consequently in the decrease of $\mu ^{(n)}$. From 
Eq.~(\ref{Eq_alpha_1}) is can be noticed that 
$\alpha (n,\chi_{l_{e}},E_{ns}) \le 1$. This term, therefore, leads to the 
increase of $\mu ^{(n)}$, which may be explained by the fact that for $n>1$ 
there are $n-1$ peaks of $\rho _{M}^{(n)}(E)$ ''generated'' by non-specific 
binding. All those peaks have smaller (mean) binding energy compared to the 
$n^{th}$ peak, which results in the smaller amount of the specifically bound 
TF. Finally, the term $\exp (-E_{ns})$ accounts for the fact that some amount 
of TF is non-specifically bound by DNA sequences, which leads to the decrease
of $\mu ^{(n)}$. The dependence of $\mu ^{(n)}$ from $E_{ns}$ is, therefore,
quite complicated, since increase in non-specific binding decreases $\mu
^{(n)}$ through $\exp (-E_{ns})$, but increases it through $\alpha (n,\chi
_{l_{e}},E_{ns})$. Similarly, the increase of $\chi _{l_{e}}$ has the
opposite effects on  $\mu ^{(n)}$ through terms $\exp [(n-1/2)\chi
_{l_{e}}^{2}\,]\,$\ and $\alpha (n,\chi _{l_{e}},E_{ns})$. 

\section{Maximum of energy distribution in SELEX with fixed selection 
stringency}

We want to determine how the position of maximum $E_{m}^{(n)}$ of 
$\rho _{M}^{(n)}(E)$ (see Eq.~(\ref{Eq_dist_unsat})), changes with the number 
of performed SELEX rounds $n$. We first shift zero of energy to coincide with 
$\overline{E}_{l_{e}}$ (see Eqs.~(\ref{Eq_Rho_M}) and~(\ref{Eq_M_mean})). 
Position of maxima of $\rho _{M}^{(n)}(E)$ is given by:
\beqar
\frac{d \rho_{M}^{(n)}(E)}{dE}\mid_{E=E_{m}^{(n)}}=0
\eeqar{Eq_Max_n_cond}
From Eq.~(\ref{Eq_dist_unsat}) and the equation above, we obtain:
\beqar
[1-f(E_{m}^{(n)}-\mu )]=-\frac{E_{m}^{(n)}}{n\,\chi _{l_{e}}^{2}}
\eeqar{Eq_Max_n}

\begin{figure}
\vspace*{7.4cm} \includegraphics{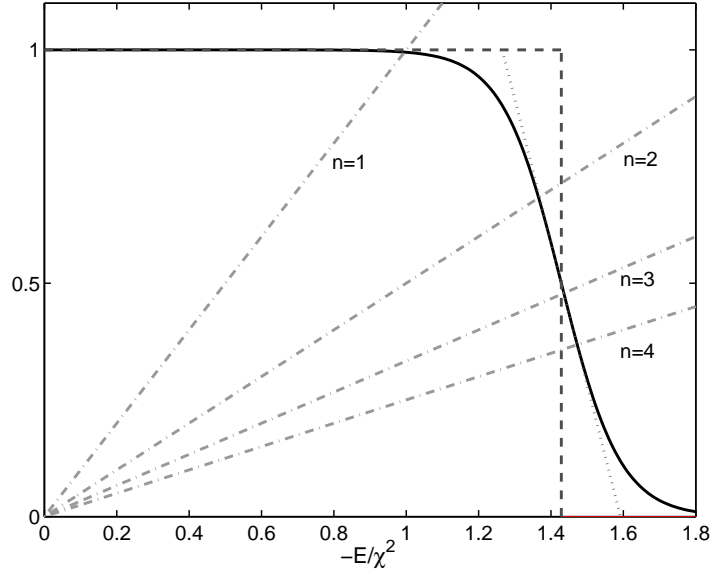} 
\caption{Full lines and dashed lines present $1-f(E-\mu )$ and unit step 
approximation to $1-f(E-\mu )$ respectively.
Dash-doted lines present family of lines $\phi _{n}\left( E\right) =-$ $%
E/(n\chi _{l_{e}}^{2})$, which is here plotted for $n=1,...,4$. The
parameter values on this figure are $\chi_{l_{e}}=3.5\,$ and 
$\mu =-5\chi_{l_{e}}$. Positions of $E_{M}^{(n)}$ are
determined by the intersections of dash-dot lines and solid curve. Note
that, for the parameters we choose, $E_{M}^{(1)}=-\chi_{l_{e}}^{2}$, while $%
E_{M}^{(n)}\approx \mu $ for $n \ge 2$.}
\label{Max_n_pos}
\end{figure}

The equation above can be solved graphically, i.e.
positions of $E_{m}^{(n)}$ (for different n) are determined by the
intersections of the family of lines $\phi _{n}\left( E\right) =-$ $E/(n\chi
_{l_{e}}^{2})$ and the curve $[1-f(E-\mu )]$. If we approximate $f(E-\mu )$
by a step $\theta (\mu -E )$ (see the dashed line in Fig.~\ref{Max_n_pos}), we 
obtain that $%
E_{m}^{(n)}=-n\,\chi _{l_{e}}^{2}$ for $n<-\mu /\chi _{l_{e}}^{2}$ and 
$E_{m}^{(n)}=\mu $ otherwise, which is shown in Fig.~\ref{Max_n_pos}. 
Correction, accurate up to the next order in $1/\chi _{l_{e}}^{2}$
can be found by linearizing $f(E-\mu )$ around $E=\mu $ (see dotted
line in Fig.~\ref{Max_n_pos}). We then obtain:
\beqar
E_{m}^{(n)}=\frac{(\mu -2)}{1+4/(\chi _{l_{e}}^{2}\,n)}
\eeqar{Eq_Max_corr}
if $n>(-\mu + 2)/\chi _{l_{e}}^{2}$, and $E_{m}^{(n)}=-n\,\chi _{l_{e}}^{2}$
otherwise. Note that, for $n$ large enough, i.e. $n \gg 4/ \chi _{l_{e}}^{2}$,
$E_m^{(n)} \rightarrow (\mu-2)$. Since in SELEX experiments $\mu $ is 
typically positioned in the
tail of the energy distribution of random binders, we expect $\mu \gg 1$
(see also the comment below Eq.~(\ref{Eq_Max_corr})), so $E_{m}^{(n)}$ for 
$n>-\mu/\chi _{l_{e}}^{2}$ is well approximated by $E_{m}^{(n)}\approx \mu $ 
(see also Fig.~\ref{Max_n_pos} and Fig.~\ref{fixed_strin_SELEX}).

\section{Computing energy matrix from extracted SELEX \\ sequences}

We determine the energy matrix $\widetilde{\epsilon }$, and the parameter $%
\gamma $ by maximizing the likelihood function $\Lambda $ (see 
Eqs.~(\ref{Eq_L}) and~(\ref{Eq_Rho_1})), subject to the constraints:
\beqar
\sum_{i,\alpha }\,p_{\alpha }\,\widetilde{\epsilon }_{i,\alpha }^{2}=1
\eeqar{Eq_quad_constr}
and
\beqar
\sum_{\alpha }\,p_{\alpha }\,\widetilde{\epsilon }_{i,\alpha }=0 \; (\forall i)
\eeqar{Eq_lin_constr}

The constraint Eq.~(\ref{Eq_quad_constr}) follows from 
$\widetilde{\epsilon }_{i,\alpha }=\epsilon
_{i,\alpha }/\chi $ and $\chi ^{2}=\sum_{i,\alpha }p_{\alpha }\epsilon
_{i,\alpha }^{2}$ (see Eq.~(\ref{Eq_chi_square})), while the constraint 
Eq.~(\ref{Eq_lin_constr}), shifts columns of 
$\widetilde{\epsilon }_{i,\alpha }$ (see the footnote 4 in 
Subsection~\ref{S4.3.2}) so 
that $\overline{\epsilon }_{i}=0$ and consequently $\overline{E}=0$ (see 
Eqs.~\ref{Eq_Rho_E} and~\ref{Eq_mean_E}). As discussed in Section~\ref{S4.3} 
and Appendix C, we approximate the binding probability 
$f(E(S)-\mu )$ in $\Lambda $ by using the strongest binding site $s_{M}$
with length $L$, on each sequence $S$ $(S\in A)$ of length $l$. To simplify
notation, we further in this Appendix use $s_{M}\equiv s$. If we use 
$\mu /\chi $ from Eq.~(\ref{Eq_constr}), variation of $\Lambda $
with respect to  $\widetilde{\epsilon }_{i,\alpha }$ and $\gamma $ leads to:
\beqar
\frac{\partial \Lambda }{\partial \widetilde{\epsilon }_{i,\alpha }}
&=&-\left( n\chi \right) \sum_{s\in A}[1-f\left( E\left( s\right) -\mu \right)
]\left( s_{i,\alpha }-s_{i,\alpha }^{\ast }\right) -\nonumber \\
&& \, - \left( n\chi \right) \gamma
\int f^{n}\left( E-\mu \right) \,\left[ 1-f\left( E-\mu \right) \right]
\,\rho _{M}\left( E\right) dE\,s_{i,\alpha }^{\ast }  -\nonumber \\
&& \,-2\alpha \,p_{\alpha }\widetilde{\epsilon }_{i,\alpha }-\lambda
\,p_{\alpha }=0,
\eeqar{Eq_epsilon}
\beqar
\frac{\partial \Lambda }{\partial \gamma }=\frac{n_{S}}{\gamma }-\int
f^{n}\left( E-\mu \right) \rho _{M}\left( E\right) dE=0,
\eeqar{Eq_gamma_1}
where $\alpha $ and $\lambda $\ are the Lagrange multipliers associated with
the constraints Eq.~(\ref{Eq_quad_constr}) and Eq.~(\ref{Eq_lin_constr}) 
respectively, while $\rho_{M}\left( E\right) $ is given by 
Eq.~(\ref{Eq_Rho_M}). Eliminating $\alpha $, $\lambda $ and $\gamma $, using 
respectively Eqs.~(\ref{Eq_quad_constr}),~(\ref{Eq_lin_constr}) 
and~(\ref{Eq_gamma_1}), leads to the equation that implicitly determines 
$\widetilde{\epsilon }_{i,\alpha }$:
\beqar
\widetilde{\epsilon }_{i,\alpha }=\frac{\frac{p_{\alpha }^{-1}}{n_{s}}%
\sum_{s\in A}[1-f\left( E\left( s\right) -\mu \right) ]\left( s_{i,\alpha
}-s_{i,\alpha }^{\ast }\right) +\left( s_{i,\alpha }^{\ast }/p_{\alpha
}-1\right) \left( 1-\nu ^{\left( n+1\right) }\right) }{\frac{1}{n_{s}}%
\sum_{s\in A}[1-f\left( E\left( s\right) -\mu \right) ]\left( E\left(
s\right) -\mu \right) +\mu \left( 1-\nu ^{\left( n+1\right) }\right) }, 
\nonumber \\
\eeqar{Eq_final}
where
\beqar
\nu ^{\left( n+1\right) }=\frac{\int f^{n+1}\left( E-\mu \right) \rho
_{M}\left( E\right) dE}{\int f^{n}\left( E-\mu \right) \rho _{M}\left(
E\right) dE}.
\eeqar{Eq_compl}

From Eq.~(\ref{Eq_Rho_n}) can be observed that $\nu ^{\left(n+1\right)}$ 
is equal to the fraction of DNA that is in complex with protein in
the $\left( n+1\right) ^{th}$ round of SELEX, when non-specific binding is
small (i.e. $n$ large enough, as discussed in Section~\ref{S4.3.1}).

In practice, we typically do not know the exact value of $\chi $, so we fix 
it to some reasonable value and then (numerically) solve Eq.~(\ref{Eq_final}) 
with respect to $\widetilde{\epsilon }_{i,\alpha }$. We can estimate the
reasonable range of $\chi $ values, if we adopt the so-called ``two state''
model~\cite{Berg_1987}, in which (a same) penalty in binding energy 
$\epsilon _{0}$ is assigned for each nucleotide that does not match the 
consensus sequence. Since one or two hydrogen bonds are formed per contact of 
the TF surface with a prefered nucleotide (energy of a hydrogen bond is 
$\sim k_{B}T$), $\epsilon _{0}$ can be estimated to be 
$(1 \sim 3) \,k_{B}T$~\cite{Gerland_2002}. From Eq.~(\ref{Eq_chi_square})
follows that $\chi \sim \sqrt{L}\epsilon _{0}$, and with $L=$ $12$ for
CTF/NFI (we are ignoring the contribution of the 3bp spacer to the binding
energy), we obtain that $\chi $ is expected to take values from 3 to 12. 
Numerical solutions of Eq.~(\ref{Eq_final}) show that for $\chi $ in this 
range, the quantity $\widetilde{\epsilon }_{i,\alpha }$ depends weakly on 
the imposed value of $\chi $. More precisely, we tested that 
$\widetilde{\epsilon }_{i,\alpha }$, which corresponds to solving 
Eq.~(\ref{Eq_final}) with (different) values of $\chi$ in the indicated range, 
leads to negligible differences in energy distribution and DET curve for 
finite T energy matrix (data now shown), which justifies solving the equation 
without the knowledge of the exact value of $\chi $. The energy matrix 
$\widetilde{\epsilon }_{i,\alpha }$ obtained by our method (corresponding to 
$\chi=5$) is given in Table~1.

\begin{table}[t]
\begin{center}
{\hspace*{-0.7cm}
{\footnotesize
\begin{tabular}{||c||c|c|c|c|c|c|c|c|c|c|c|c|c|c|c||} \hline \hline
& 1& 2& 3& 4& 5& 6& 7& 8& 9& 10& 11& 12& 13& 14& 15\\ \hline \hline
A & 0.13& 0.23& 0.24& 0.13& -0.07& -0.16& 0.03& 0.02& 0.00& 0.05& 0.18& 
0.19&
0.24& -0.54& -0.31 \\ \hline
T & -0.29& -0.54& 0.24& 0.23& 0.18& 0.01& -0.02& 0.02& 0.03& -0.17& -0.07&
0.21& 0.24& 0.22& 0.15 \\ \hline
C & -0.04& 0.20& 0.24& 0.24& -0.32& 0.07& -0.03& 0.00& 0.03& 0.07& 0.21&
-0.64& -0.72& 0.13& 0.21 \\ \hline
G & 0.20& 0.11& -0.72& -0.61& 0.21& 0.09& 0.02& -0.04& -0.06& 0.04& -0.33&
0.24& 0.24& 0.19& -0.05 \\ \hline \hline
\end{tabular}
\vskip 4truemm 
\hspace*{-0.7cm}
\begin{tabular}{||c||c|c|c|c|c|c|c|c|c|c|c|c|c|c|c||} \hline \hline
& 1& 2& 3& 4& 5& 6& 7& 8& 9& 10& 11& 12& 13& 14& 15\\ \hline \hline
A & 0.07& 0.28& 0.26& -0.15& -0.15& -0.11& 0.04& 0.04& 0.01& 0.01& 0.18& 
-0.02&
0.26& -0.39& -0.23 \\ \hline
T & -0.22& -0.40& 0.26& 0.23& 0.21& 0.02& -0.03& 0.02& -0.01& -0.11& -0.12&
0.08& 0.26& 0.29& 0.07 \\ \hline
C & -0.05& 0.18& 0.26& 0.49& -0.25& 0.05& -0.01& -0.01& 0.03& 0.07& 0.21&
-0.55& -0.79& -0.04& 0.26 \\ \hline
G & 0.20& -0.05& -0.79& -0.56& 0.19& 0.04& 0.00& -0.04& -0.03& 0.03& -0.27&
0.50& 0.26& 0.15& -0.10 \\ \hline \hline
\end{tabular}
\vskip 4truemm 
\hspace*{-0.7cm}
\begin{tabular}{||c||c|c|c|c|c|c|c|c|c|c|c|c|c|c|c||} \hline \hline
& 1& 2& 3& 4& 5& 6& 7& 8& 9& 10& 11& 12& 13& 14& 15\\ \hline \hline
A & 0.06& 0.19& 0.29& 0.16& -0.03& -0.04& 0.01& 0.01& 0.00& 0.01& 0.11& 
0.21&
0.29& -0.48& -0.16 \\ \hline
T & -0.16& -0.50& 0.29& 0.23& 0.11& 0.01& -0.01& 0.01& 0.00& -0.05& -0.01&
0.23& 0.29& 0.19& 0.07 \\ \hline
C & 0.01& 0.18& 0.29& 0.23& -0.18& 0.02& 0.00& 0.00& 0.01& 0.02& 0.11&
-0.69& -0.86& 0.12& 0.10 \\ \hline
G & 0.09& 0.12& -0.86& -0.63& 0.10& 0.02& 0.00& -0.01& -0.01& 0.02& -0.21&
0.25& 0.29& 0.17& -0.01 \\ \hline \hline
\end{tabular}}}
\end{center}
\caption{ Upper table: The finite T energy matrix for CTF/NFI transcription 
factor. Middle table: The information-theoretic weight matrix. 
Lower table: The MatInspector weight matrix.}
\end{table}
              
\begin{figure}
\vspace*{5.9cm} \includegraphics{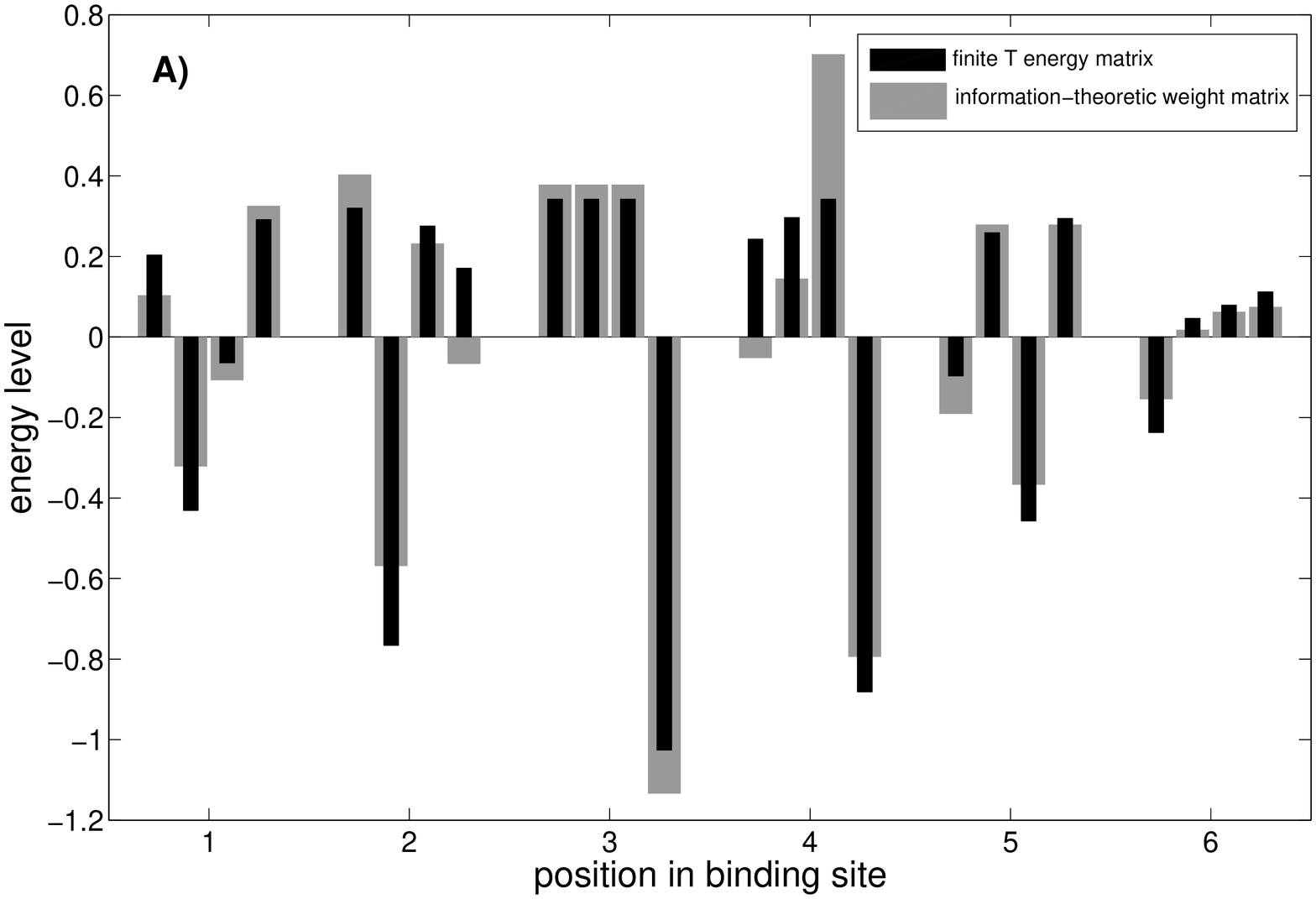} 
\vspace*{6.9cm} \includegraphics{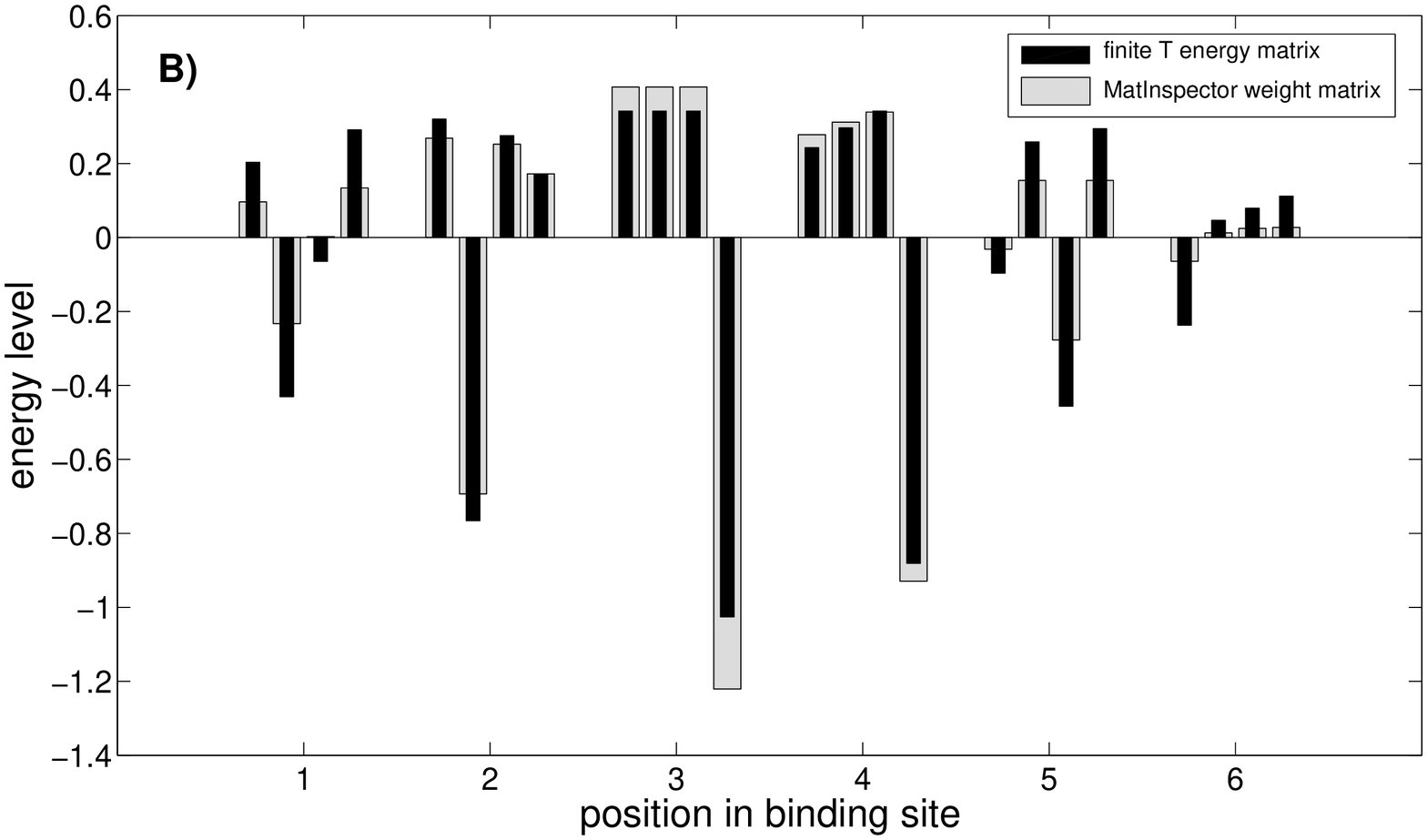} 
\caption{Comparison of the finite T matrix elements with A) the information 
theory weight matrix and B) the MatInspector matrix elements. CTF/NFI binds 
DNA as a homodimer, and recognizes two $6$ bp long palindrome symmetric 
motifs, separated by a $3$ bp spacer. Consequently, matrices in Table 1 were 
appropriately symmetrized and only the first six positions are shown. Four 
bars at each position on the horizontal axis correspond to A, T, C and G 
respectively, while heights of the bars correspond to the values of the matrix 
elements.}
\label{matrices_comparison}
\end{figure}

Finally, we use Table 1 to discuss the differences between the weight 
matrix parameters estimated by the three different methods. In 
Fig.~\ref{matrices_comparison}A histograms of energy levels corresponding to 
the finite T energy matrix and to the information-theoretic weight matrix are  
shown together. Similarly Fig.~\ref{matrices_comparison}B compares the matrix 
parameters corresponding to the finite T energy matrix and the MatInspector 
weight matrix.  The figure directly indicates which parameters are most 
different between the two matrices. For example, there is $\sim 150\%$ 
difference between the information-theoretic and our matrix at the position 
2G, $\sim 100\%$ difference at the positions 4C and 5A, etc. Similarly, 
comparison with MatInspector weight matrix in Fig.~\ref{matrices_comparison}B 
shows that there are significant differences (from $40\%$ to $100 \%$) between 
the two matrices at the positions 1, 5 and 6 while the matrices mostly agree 
with each other at the positions 2,3 and 4. Since positions 2, 3 and 4 
contribute more to the binding energy than positions 1, 5 and 6, the smaller 
difference between the DET curves of the finite T and the MatInspector 
matrices (as compared to the difference between the finite T and the 
information theory DET curves) can be attributed to the localization of the 
matrix differences at the less conserved positions. We note that, while the 
differences between the individual matrix elements are generally not very 
large, when binding site scores are calculated the individual differences add 
up to produce significant differences in e.g. the false positive/false 
negative trade-off shown in Fig.~\ref{DET_curve}.

\newpage

\end{appendix}

\end{document}